\documentclass[
aps,
pre,
reprint,
amsmath,
amssymb,
floatfix
]{revtex4-2}

\usepackage{graphicx}
\usepackage{dcolumn}
\usepackage{bm}
\usepackage{url}

\begin{document}

\title{Multidimensional dynamical centrality from Green functions in complex networks}

\author{Yusen Wang$^{1, 2}$}
\author{Hao Yu$^{1}$}
\email{Contact author: Hao.Yu@xjtlu.edu.cn}
\affiliation{$^1$Department of Physics, Xi'an Jiaotong-Liverpool University, Suzhou, China}
\affiliation{$^2$Department of Physics, University College London, London, United Kingdom}

\date{\today}

\begin{abstract}
	Conventional centrality measures provide compact descriptions of node importance, but they emphasize specific structural relations and do not directly resolve the temporal and spectral organization of dynamical perturbation responses. Building on the Green function of a linearized networked system, we develop a multidimensional framework for dynamical node characterization. From the same response function, we extract three complementary indicators---Influence, Efficiency, and Distortion---that quantify cumulative response strength, temporal rate of response, and spectral concentration, respectively, together with a contribution matrix that resolves these quantities into source--target pathways. As a numerical demonstration, we apply the framework to weighted Kuramoto--Sakaguchi dynamics on heterogeneous Barab\'asi--Albert networks. The three indicators distinguish implanted dynamical node classes through complementary dimensions of the perturbation response, while remaining strongly coupled to conventional topological centralities. These results demonstrate how a common dynamical response can be decomposed into distinct, physically interpretable characteristics rather than represented by a single aggregate descriptor. We finally discuss the scope and limitations of the linear-response assumption and extensions to time-dependent and nonlinear regimes.
\end{abstract}

\maketitle

\section{Introduction}

The identification of important nodes in complex networks is a fundamental problem in network science, with broad implications for epidemic containment~\cite{albert2000,pastor2001,cohen2001}, power-grid stability, neural information routing, and social influence maximization. In these systems, the functional importance of a node is shaped not only by its topological location, but also by how perturbations propagate through the underlying dynamics.

Conventional approaches quantify node importance through topological centrality measures such as degree, betweenness, closeness, and eigenvector centrality~\cite{freeman1978,bonacich1987,newman2010}. These measures characterize structural position and have proved useful for identifying hubs, bottlenecks, and influential locations. Centrality is not, however, a unique concept: different measures encode different assumptions about how importance is related to network structure and how interactions are transmitted through the network~\cite{freeman1978,oldham2019}. A comparative analysis of seventeen centrality measures across two hundred twelve real-world networks has shown that centrality measures can provide distinct information about nodal roles, with their mutual relationships depending on the underlying topology~\cite{oldham2019}. Thus, even within a purely structural description, node importance is intrinsically multidimensional rather than uniquely specified by a single centrality coordinate.

More importantly, structural position does not uniquely determine dynamical importance. The effect of a node on collective behavior depends on the interplay between network structure and the dynamics operating on that structure~\cite{Klemm2012}. Perturbative analyses of information flow have likewise shown that the same network topology can support qualitatively different propagation patterns under different interaction dynamics~\cite{Harush2017}. Causal-intervention studies in specific nonequilibrium systems have further found that static structural features can be poor predictors of dynamic node importance~\cite{Elteren2022}. These results indicate that structural centrality should not be interpreted as a complete description of a node's functional role once the dynamics of propagation are taken into account.

Motivated by the distinction between dynamics and structure, dynamical notions of node importance have been developed from several perspectives. One direction quantifies how modifying a node or link changes a global dynamical property. For example, Restrepo, Ott, and Hunt defined dynamical importance through the relative change of the leading eigenvalue following node or link removal~\cite{restrepo2006}. Other approaches include network controllability, which characterizes the ability of selected nodes to steer a network~\cite{liu2011}, cascading-failure analyses~\cite{motter2002}, and influence maximization under specific spreading rules~\cite{kitsak2010}. Such methods incorporate dynamical considerations beyond a purely structural description, but the resulting node scores are generally tied to a particular dynamical criterion, perturbation protocol, or network modification.

Another related line of work characterizes dynamical interactions through matrix exponentials. Estrada and Hatano introduced communicability and subgraph centrality based on $e^{\mathbf{A}}$ and its diagonal elements~\cite{estrada2005,estrada2008}. Gilson and co-workers subsequently introduced a dynamic communicability framework in which the Green function $G(t)=e^{\mathbf{J}t}$ of a linearized dynamical system provides a time-resolved description of interactions between nodes~\cite{gilson2018framework}. Related formulations have also been applied to fMRI transition dynamics during movie viewing~\cite{gilson2018neuro} and reciprocal influence in trade networks~\cite{bartesaghi2022}. These studies establish the Green function as an effective representation of dynamical interactions, yet they typically reduce the response to a single communicability scalar.

Here we build on this response representation to systematically resolve a single response operator into several physically interpretable dimensions of node importance. From the same Green function, we define three indicators---Influence, Efficiency, and Distortion---that characterize, respectively, the cumulative strength, temporal organization, and spectral organization of the propagated response. This construction separates different aspects of propagation without requiring a different dynamical model or response object for each indicator. In addition, we introduce a normalized contribution matrix that traces the aggregate response back to individual source--target pairs and temporal or spectral components, providing a pathway-level interpretation of the node characteristics.

The framework is demonstrated using weighted Kuramoto--Sakaguchi dynamics on heterogeneous Barab\'asi--Albert networks. We examine how the indicators relate to conventional structural centralities, how consistently they distinguish nodes with controlled local dynamical modifications across network realizations, and how the contribution representation identifies the pathways underlying the aggregate response. The comparison is designed to assess how strongly dynamical response is coupled to structural position, and what additional physical organization is retained in the response itself.

The remainder of the paper develops the linear-response construction, defines the three indicators and their contribution representation, and examines their behavior in numerical network realizations. We finally discuss the conditions under which the linear Green function description is applicable, together with limitations associated with strong perturbations, near-marginal stability, and time-dependent dynamics.
\section{Method}

\subsection{General linear response formalism}

Linear response theory provides a general framework for describing the response of a dynamical system to a weak external perturbation. For an observable $A$, its first-order response to a small stimulus $u(t)$ takes the convolution form
\begin{equation}
    \delta \langle A(t) \rangle = \int_{-\infty}^{t} R(t-\tau) u(\tau)\, d\tau,
\end{equation}
where $R(t)$ is the response kernel, which describes the causal relationship between stimulus and response. This form was originally established by Kubo for equilibrium statistical mechanics~\cite{Kubo1957,Kubo1966}, and has become a standard tool in linear-response and control theory for both equilibrium and non-equilibrium systems.

For general differentiable dynamical systems, Ruelle extended linear response theory to non-Hamiltonian settings~\cite{Ruelle1998,Ruelle2009}. In the special case considered here, where the unperturbed dynamics has a stable reference state, the response can be represented by the tangent dynamics obtained from the local linearization of the flow.

\subsection{Specialization to a stable reference state}

Consider a deterministic system of the form
\begin{equation}
	\dot{\bm{\Gamma}}
	=
	\mathbf{F}(\bm{\Gamma})
	+
	\mathbf{B}\mathbf{u}(t),
\end{equation}
where $\bm{\Gamma}\in\mathbb{R}^N$ is the state vector,
$\mathbf{F}$ is a smooth vector field,
$\mathbf{B}\in\mathbb{R}^{N\times p}$ is the input matrix, and
$\mathbf{u}(t)\in\mathbb{R}^p$ is a small stimulus. The evolution of an infinitesimal perturbation $\delta\bm{\Gamma}$ is
governed by the variational equation
\begin{equation}
	\delta\dot{\bm{\Gamma}}
	=
	\mathbf{DF}(\bm{\Gamma}_s)\,\delta\bm{\Gamma}
	+
	\mathbf{B}\mathbf{u}(t),
\end{equation}
where $\mathbf{DF}(\bm{\Gamma}_s)$ denotes the Jacobian evaluated at the reference state.

We now impose the key specialization: assume that the unperturbed
system has a stable reference state \(\bm{\Gamma}_s\), which may correspond to a fixed point or a phase-locked state represented as a stationary configuration in an appropriate co-rotating frame, and restrict the analysis to the linear
neighborhood of this state. Writing
$\bm{\Gamma}(t)=\bm{\Gamma}_s+\delta\bm{\Gamma}(t)$ and linearizing about $\bm{\Gamma}_s$, we then obtain
\begin{equation}
	\delta\dot{\bm{\Gamma}}
	=
	\mathbf{J}\,\delta\bm{\Gamma}
	+
	\mathbf{B}\mathbf{u}(t),
	\label{eq:ODE}
\end{equation}
with the Jacobian matrix
\begin{equation}
	\mathbf{J}
	=
	\mathbf{DF}(\bm{\Gamma}_s)
	=
	\left.
	\frac{\partial F_i}{\partial \Gamma_j}
	\right|_{\bm{\Gamma}_s}.
\end{equation}

Solving Eq.~(\ref{eq:ODE}) with the initial condition
$\delta\bm{\Gamma}(0)=0$ via the variation-of-constants formula, we
obtain
\begin{equation}
	\delta\bm{\Gamma}(t)
	=
	\int_0^t
	e^{\mathbf{J}(t-\tau)}
	\mathbf{B}\mathbf{u}(\tau)\,d\tau.
	\label{eq:variation}
\end{equation}

Next, we define the matrix valued Green function
\begin{equation}
	\mathbf{G}(t)
	:=
	e^{\mathbf{J}t}
	\quad (t\geq0),
	\qquad
	\mathbf{G}(t)=0
	\quad (t<0),
\end{equation}
so Eq.~(\ref{eq:variation}) can be written as
\begin{equation}
	\delta\bm{\Gamma}
	=
	\mathbf{G}*(\mathbf{B}\mathbf{u}).
\end{equation}

For a scalar observable $A(\bm{\Gamma})$, the first-order response is
\begin{equation}
	\delta A(t)
	\simeq
	\nabla A(\bm{\Gamma}_s)^{\mathsf T}
	\delta\bm{\Gamma}(t)
	=
	\int_0^t
	\nabla A(\bm{\Gamma}_s)^{\mathsf T}
	\mathbf{G}(t-\tau)
	\mathbf{B}\mathbf{u}(\tau)\,d\tau.
\end{equation}
Thus, the corresponding response kernel is
\begin{equation}
	R_A(t)
	=
	\nabla A(\bm{\Gamma}_s)^{\mathsf T}
	\mathbf{G}(t)
	\mathbf{B}.
\end{equation}

For the full-state response, the observable is $A(\bm{\Gamma})=\bm{\Gamma}$, whose Jacobian with respect to the state is the identity matrix. In the particular case where the input acts on all $N$ state components independently, we have $p=N$, $\mathbf{B}=\mathbb{I}$, and the response operator reduces to
\begin{equation}
	\mathbf{R}(t)=\mathbf{G}(t).
\end{equation}
Thus, in the full-state, componentwise-input setting considered here,
the Green function of the linearized dynamics provides the
matrix-valued linear response kernel.

This construction is closely related to the dynamic communicability
framework proposed by Gilson et al.~\cite{gilson2018framework}, which
also uses $G(t)=e^{\mathbf{J}t}$ to characterize time-resolved
interactions. The present work develops a multidimensional
analysis and contribution-resolved framework based on the same
response object.

\subsection{Green functions on complex networks}

The formalism applies directly to complex dynamical networks, where $\Gamma_i$ denotes the state of node $i$ in an $N$-node network. The network topology and coupling enter through $\mathbf{F}$, and the Jacobian $\mathbf{J}$ describes the effective linearized interactions. The matrix element $G_{ij}(t)$ has a clear physical interpretation: it is the response of node $i$ at time $t$ to a unit impulse applied to node $j$ at time $0$.

For the Kuramoto--Sakaguchi dynamics, a global phase shift generates
a neutral mode. To suppress this symmetry from the propagator, we
project the dynamics onto the subspace orthogonal to the uniform phase direction. Let $\mathbf{U}$ be an orthonormal basis of this subspace, satisfying $\mathbf{U}^{\mathsf T}\mathbf{1}=0$. We then define the projected Jacobian
\begin{equation}
	\mathbf{J}_{\mathrm{red}}=\mathbf{U}^{\mathsf T}\mathbf{J}\mathbf{U},
\end{equation}
and evaluate the phase-difference Green function as
\begin{equation}
	\mathbf{G}(t)=\mathbf{U}e^{\mathbf{J}_{\mathrm{red}}t}\mathbf{U}^{\mathsf T}.
\end{equation}
This projection suppresses the global phase component while preserving the original $N\times N$ node representation. For the parameter regime considered here, the projected Jacobian has no eigenvalue with positive real part, and the resulting Green function decays to zero at long times.

\subsection{Dynamical centrality indicators}

The dynamical centrality of a node is characterized by three complementary indicators derived from the same Green function response.

Consider a source node $s$. The total response intensity transmitted from $s$ to the rest of the network is defined as
\begin{equation}
    R_s(t) = \sum_{i \neq s} |G_{is}(t)|^2,
\end{equation}
where the self-term $i=s$ is excluded because it contains the instantaneous response and does not represent propagation. The squared modulus $|G_{is}(t)|^2$ is used because it provides a non-negative measure of the instantaneous response intensity. For real-valued dynamics, it coincides with the squared response amplitude; for complex-valued modes, it removes the phase factor, which is not directly relevant to the magnitude. Although this operation discards phase information, it is appropriate for the present framework because the three indicators $I_s$, $E_s$, and $D_s$ are designed to capture magnitude-related aspects of the response. Phase information can be retained separately if one wishes to study synchronization properties or wave-like propagation, but it is not required for the centrality characterization developed here.

\paragraph{Influence}
$I_s$ quantifies the cumulative response strength transmitted from node $s$ to the rest of the network:
\begin{equation}
    I_s = \frac{1}{N-1} \int_0^T \sum_{i \neq s} |G_{is}(t)|^2\, dt.
    \label{eq:influence}
\end{equation}
The upper integration limit $T$ is chosen as a finite observation
window rather than infinity, such that the global response intensity
has decayed to a negligible level by the end of the integration
interval. Specifically, we require
\begin{equation}
	\max_s\frac{R_s(T)}{\max_t R_s(t)}<10^{-3}.
\end{equation}
This criterion provides a practical truncation rule for the numerical
evaluation of the response integrals. For the simulations presented
here, $T=50$ satisfies this criterion for all nodes; the detailed
numerical verification is given in Appendix~\ref{sec:truncation}

This indicator describes the cumulative magnitude of the response propagated from the source node $s$ across the network, integrating contributions over both target nodes and time.

\paragraph{Efficiency}
$E_s$ quantifies the temporal rate of the perturbation response. We first define the intensity-weighted mean response time
\begin{equation}
    \tau_s = \frac{\int_0^T t\, R_s(t)\, dt}{\int_0^T R_s(t)\, dt},
\end{equation}
where the response intensity $R_s(t)$ serves as the weighting factor. The efficiency is then defined as
\begin{equation}
    E_s = \frac{1}{\tau_s}.
\end{equation}
Thus, a larger $E_s$ corresponds to a shorter characteristic response time.

\paragraph{Distortion}
$D_s$ quantifies the spectral distortion of the transmitted response. 
The spectral flatness of the path $s\to i$, expressed through the participation ratio~\cite{Thouless1974,Wegner1980}, is defined as
\begin{equation}
    \mathrm{SF}_{is} = \frac{\left(\sum_k P_{is}(\omega_k)\right)^2}{\sum_k P_{is}(\omega_k)^2}.
\end{equation}
The participation ratio, originally introduced to characterize the spatial extension of vibrational modes in disordered systems~\cite{BellDean1970}, measures the effective number of frequency components contributing to the response. If the spectrum is perfectly flat, all bins contribute equally and $\mathrm{SF}_{is}$ is equal to the total number of retained bins; if the spectrum collapses to a single dominant frequency, $\mathrm{SF}_{is}$ approaches 1. Thus, flatness refers to the uniformity of the power distribution across frequency bins: a flat spectrum distributes energy evenly among retained frequencies, with no single oscillatory mode preferentially excited; a concentrated spectrum, by contrast, indicates that most of the energy is distributed in a few dominant modes. Here the quantity is used to characterize spectral concentration of the propagated response rather than as a direct measure of signal fidelity.

The power spectrum contains a broad low-frequency component associated with the decay envelope of the Green function. This component primarily reflects the global relaxation of the linearized system rather than specific path propagation structure. To reduce its effect on the spectral analysis, we first remove the temporal mean of $|G_{is}(t)|^2$, which eliminates the zero-frequency (DC) component. The power spectrum is defined as
\begin{equation}
    P_{is}(\omega_k)=\left|\mathcal{F}\left[|G_{is}(t)|^2-\left\langle |G_{is}(t)|^2\right\rangle_t\right](\omega_k)\right|^2,
\end{equation}
where $\omega_k$ denotes the discrete frequency associated with the $k$th Fourier bin. To further suppress the remaining low-frequency contribution, we exclude the low-frequency bins whose period is larger than the characteristic decay time of the slowest mode. In the present simulations, the frequency resolution is approximately $0.02$ Hz, and we retain frequencies above $0.6$ Hz, corresponding to discarding the first $30$ spectral bins. This threshold lies in the separation between the broad low-frequency decay envelope and the higher-frequency oscillations. This procedure is analogous to standard detrending or high-pass filtering in signal processing~\cite{bendat2010,percival1993}, in which background drifts are removed so that the fluctuating components of interest can be isolated. The removal of these low-frequency components is intended to reduce the influence of the shared relaxation envelope and to highlight the oscillatory structure used in the present definition of $D_s$.

After normalization by the number of retained frequency bins, the spectral flatness of each path is normalized so that the ratio $\mathrm{SF}_{is}/M_{\text{used}}$ equals $1$ for a perfectly flat spectrum and reaches its minimum value $1/M_{\text{used}}$ when the spectrum is concentrated in a single retained frequency bin. The distortion of node $s$ is then defined as
\begin{equation}
    D_s = 1 - \frac{1}{N-1} \sum_{i \neq s} \frac{\mathrm{SF}_{is}}{M_{\text{used}}},
\end{equation}
where $M_{\text{used}}$ is the number of retained frequency bins. Therefore, for a finite number of retained frequency bins, the distortion lies in $[0,1-1/M_{\text{used}}]$, with lower values indicating a broader and less spectrally concentrated response and higher values indicating stronger spectral concentration.

A low $D_s$ indicates a spectrally flat transmission, whereas a high $D_s$ reflects strong frequency selectivity and spectral concentration. This interpretation has direct practical consequences. In power grids, a perturbation injected from a low-$D_s$ bus propagates without exciting dominant oscillatory modes, whereas a high-$D_s$ bus selectively excites a few inter-area oscillation modes, potentially producing long-lived, spectrally concentrated transients that are harder to damp. In neural circuits, low-$D_s$ neurons may exhibit broader spectral transmission, whereas high-$D_s$ neurons may preferentially emphasize a smaller number of frequency components. In infrastructure networks, high-$D_s$ nodes may correspond to locations where a disturbance is converted into a narrowband resonance, complicating detection and control.

\subsection{Contribution matrix}

Building on the Green function response framework described above, we introduce a contribution matrix that decomposes the indicators into contributions from individual target nodes and propagation paths:
\begin{equation}
    C_{is}(t) = \frac{|G_{is}(t)|^2}{(N-1)I_s}.
    \label{eq:con}
\end{equation}
This matrix decomposes the total influence $I_s$ of source node $s$ into time contributions arriving at each target node $i$. The same object serves as the basis for decomposing the other two indicators.

For Efficiency, the contribution of each spatiotemporal component is obtained by weighting $C_{is}(t)$ by the arrival time,
\begin{equation}
    C_{is}^{E}(t) = t\, C_{is}(t),
\end{equation}
so that the weighted mean arrival time can be expressed as
\begin{equation}
    \tau_s = \sum_{i\neq s}\int_0^T t\, C_{is}(t)\, dt.
\end{equation}
Regions with large $C_{is}^{E}(t)$ represent the target nodes and time intervals that contribute most strongly to the temporal weighting of the response.

For Distortion, the path-resolved quantity $C_{is}^{D}$ measures the contribution of the corresponding path to the normalized spectral-flatness component entering $D_s$: the Fourier transform of each resolved response component gives the power spectrum of a specific path $P_{is}(\omega_k)$, from which the spectral flatness $\mathrm{SF}_{is}$ is obtained. The contribution of path $s \to i$ to the distortion of node $s$ is then
\begin{equation}
    C_{is}^{D} = \frac{1}{N-1} \frac{\mathrm{SF}_{is}}{M_{\text{used}}},
\end{equation}
so that
\begin{equation}
    D_s = 1 - \sum_{i \neq s} C_{is}^{D}.
\end{equation}
This decomposition identifies which target paths contribute most
strongly to the spectral-flatness term entering $D_s$; the corresponding frequency dependence is resolved by $P_{is}(\omega_k)$.

Thus, $C_{is}(t)$ allows the tracing of all three indicators. For any source node $s$, one can identify which target nodes, in which time intervals, and in which frequency bands dominate each aspect of the dynamical centrality. This representation provides traceable information that enables the identification of the dominant sources underlying each indicator and diagnostic.

\subsection{Numerical demonstration and parameter settings}

We simulate the weighted Kuramoto--Sakaguchi model on Barab\'asi--Albert (BA)~\cite{acebron2005} networks with $N=200$ nodes and attachment parameter $m=4$~\cite{barabasi1999}. The choice $N=200$ corresponds to a mesoscale size commonly encountered in empirical systems, including whole-brain parcellations~\cite{Craddock2012,Schaefer2018}, international trade networks~\cite{clemente2023}, and power-grid network models~\cite{Pagani2014}, and it provides adequate statistical resolution for the correlation analyses reported below: For $n=200$, the two-sided Pearson-correlation test at the $0.05$ level corresponds to a threshold of approximately $|r|=0.14$; the exact inferential precision depends on the effect size and confidence interval~\cite{Bonett2000}. The computational load is a further consideration. Although the Jacobian $\mathbf{J}$ is sparse for a sparse network, the matrix exponential $G(t)=e^{\mathbf{J}t}$ is generally dense, and direct evaluation of the full matrix exponential has cubic computational scaling and quadratic storage requirements~\cite{Higham2008}; for large networks, Krylov-subspace and related matrix-function methods can be used to avoid forming the full matrix explicitly~\cite{AlMohy2011,BenziBoito2020}. The choice $N=200$ thus allows the full Green function structure to be resolved explicitly while keeping the computational load tractable.

The attachment parameter is set to $m=4$, giving an asymptotic mean
degree $\langle k\rangle\simeq2m\simeq8$. This produces a sparse but
sufficiently interconnected heterogeneous network with a clear
contrast between hubs and peripheral nodes. The connectivity scale is
consistent with related network-dynamics studies: networked Kuramoto
simulations have used $N=200$ BA networks with mean degree close to 10 ~\cite{Moreira2019}, and other studies have used $N=200$ BA networks with smaller attachment parameters~\cite{Novelli2021}. The values $N=200$ and $m=4$ are thus not assumed to be universal, but define a regime that is sufficiently large and heterogeneous for meaningful comparisons while remaining computationally tractable.

The dynamics reads
\begin{equation}
	\dot{\theta}_i=\frac{\omega_i+K\sum_j W_{ij}\sin(\theta_j-\theta_i-\alpha)}{1+\gamma_i},
\end{equation}
where $\omega_i\sim\mathcal{N}(0,1)$ are natural frequencies,
$W_{ij}$ are coupling weights associated with the network edges, $K$ is coupling strength, and
$\alpha$ is the phase-lag parameter of the Sakaguchi--Kuramoto
interaction~\cite{sakaguchi1986}. In particular, $W_{ij}=0$ for
nonexistent network connections, while the weight assignment is kept
fixed for a given network realization. The factor $1+\gamma_i$
represents heterogeneous local response timescales. Defining
$\beta_i=1+\gamma_i$, the local phase evolves on a timescale
proportional to $\beta_i$, with larger $\gamma_i$ corresponding to
slower local response. The phase lag is set to $\alpha=0.1$ rad, and
initial phases are uniformly distributed in $[0,2\pi)$ (see Appendix~\ref{sec:parameters} for full statistics). The system evolves until the order parameter $r(t)$ reaches a
stationary value, as determined by the criterion
$|r(t) - r(t - \Delta t_{\mathrm{check}})| < 10^{-6}$
for $20$ consecutive checks, where $\Delta t_{\mathrm{check}} = 1.0$
is the interval between successive evaluations of $r$. This criterion
ensures that the macroscopic degree of synchronization no longer
changes appreciably over the observation window.

For this dynamics, the Jacobian matrix $\mathbf{J}$ evaluated at the
reference state $\bm{\theta}^*$ has the following structure. For
$i \neq j$,
\begin{equation}
	J_{ij}=\frac{K W_{ij}\cos(\theta_j^* - \theta_i^* - \alpha)}{1 + \gamma_i},
\end{equation}
while the diagonal elements are defined by the requirement that each
row sums to zero,
\begin{equation}
	J_{ii}=	-\sum_{j \neq i} J_{ij}.
\end{equation}
The factor $1 + \gamma_i$ enters only through the row index $i$, so
the Jacobian is generally non-symmetric. 

From this stationary synchronized state, represented in the
co-rotating frame, we compute the Jacobian matrix $\mathbf{J}$ of the
vector field and construct the matrix Green function
$\mathbf{G}(t)=e^{\mathbf{J}t}$. The three centrality indicators and
the contribution matrix are then evaluated from $G_{ij}(t)$.

To systematically test the sensitivity of the proposed indicators to
dynamical heterogeneity, we designate two groups of nodes according to their degree. Nodes with the highest degrees are designated as
superbroadcasters and assigned enhanced coupling and weaker damping, whereas nodes with the lowest degrees are designated as slow nodes and assigned reduced coupling and stronger damping. The specific modifications are applied while keeping the underlying network
topology unchanged. This construction provides two controlled node
classes with distinct local dynamical characteristics for testing the
response-based indicators.

In heterogeneous networks, hubs can participate preferentially in the
dynamically coherent core, although the precise relationship between
degree and synchronization depends on the coupling and frequency
distributions~\cite{zhou2006}. By modifying high-degree and low-degree nodes separately, we test how the proposed indicators respond to local dynamical modifications imposed at structurally distinct network positions. The heterogeneous coupling weights $W_{ij}$ further allow the response to depend on both network connectivity and local coupling strength. This setup therefore provides a controlled comparison between structural position and local dynamical modification while leaving the underlying network topology unchanged.

\section{Results}

\subsection{Dynamical centrality compared to structural centrality}

\begin{figure}[t]
\centering
\includegraphics[width=\columnwidth]{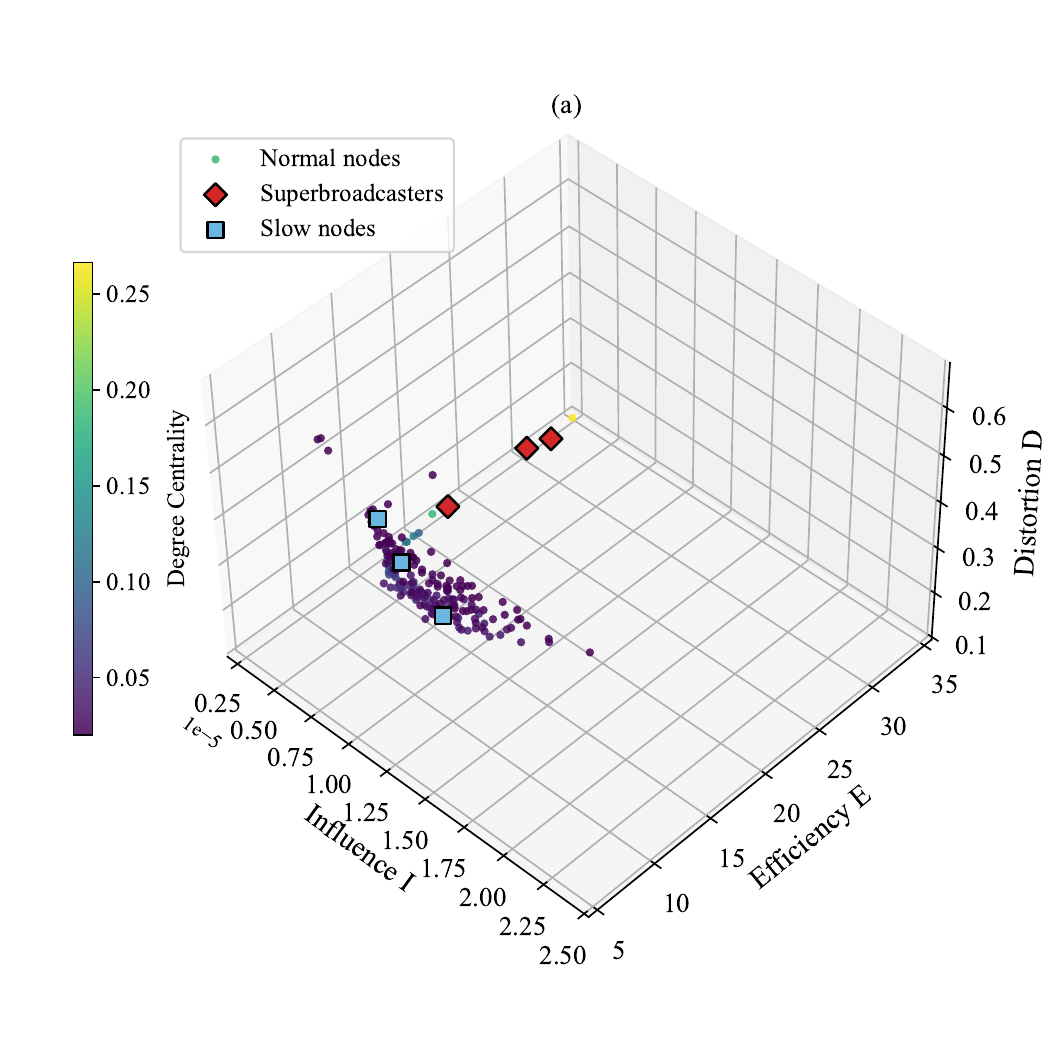}
\includegraphics[width=\columnwidth]{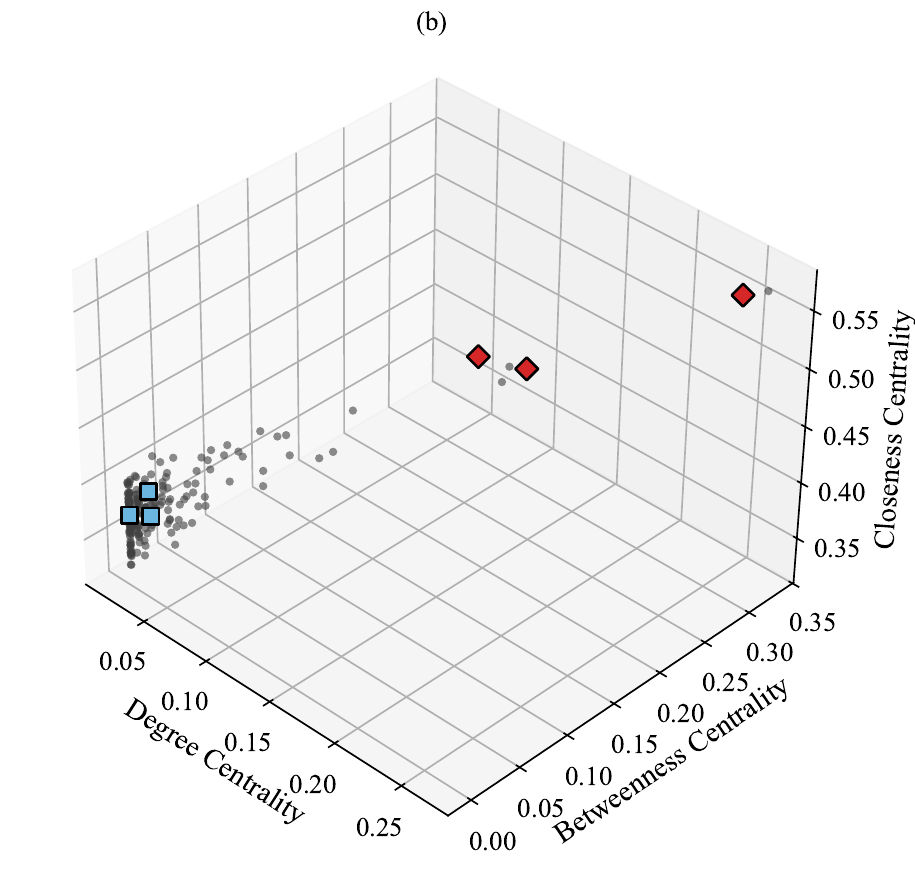}
\caption{
Comparison between dynamical and topological centrality spaces.
(a) Distribution of nodes in the dynamical centrality space defined by Influence $I$, Efficiency $E$, and Distortion $D$. Normal nodes are colored according to their degree centrality, while superbroadcasters and slow nodes are highlighted by red diamonds and blue squares, respectively.
(b) Distribution of the same nodes in the conventional topological centrality space defined by degree, betweenness, and closeness centralities.
}
\label{fig:centrality_spaces}
\end{figure}

We first examine how nodes are distributed in the $I, E, D$ three-dimensional space. Figure~\ref{fig:centrality_spaces}(a) shows the dynamical centrality space for a simulated Barab\'asi--Albert network, with nodes colored according to their degree centrality. The three axes resolve the nodes into a broad, three-dimensional cloud, and the implanted superbroadcasters and slow nodes occupy distinct regions of the dynamical space, with the clearest separation along the Efficiency and Distortion axes. The separation is not simply a consequence of introducing an additional structural coordinate, because the dynamical measures quantify temporal and spectral properties of the response in addition to its cumulative magnitude.

The same nodes are projected onto the topological space of degree, betweenness, and closeness, shown in Fig.~\ref{fig:centrality_spaces}(b). In this realization, the nodes occupy a strongly compressed region of the structural centrality space, consistent with the substantial correlations among these measures~\cite{oldham2019}. Topological measures already organize a substantial fraction of node variation, but the dynamical space resolves this variation in physically distinct response dimensions.

\begin{figure}[htbp]
\centering
\includegraphics[width=0.5\textwidth]{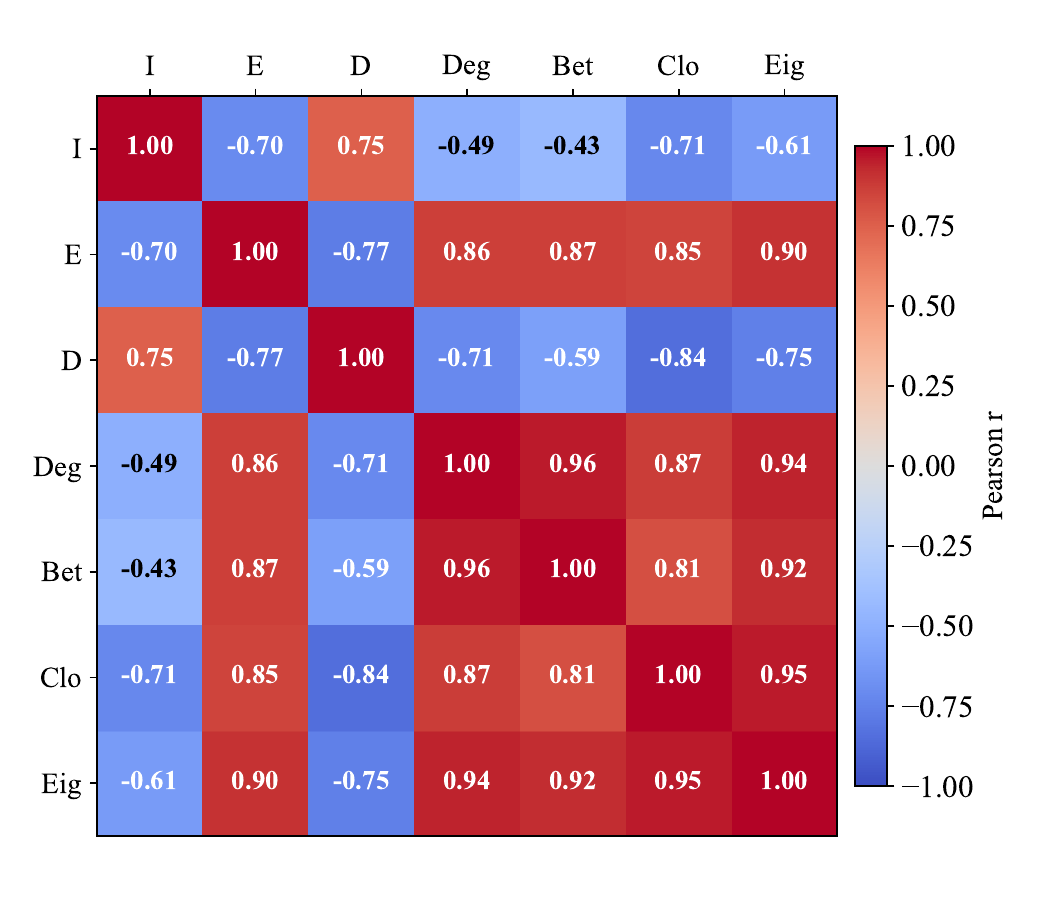}
\caption{
Pearson correlation matrix among the three dynamical indicators and four topological centrality measures.
}
\label{fig:correlation}
\end{figure}

To further quantify the relationship between the dynamical and
structural descriptions, we compute the Pearson correlation matrix among $I$, $E$, $D$, and the four topological measures, shown in
Fig.~\ref{fig:correlation}. The topological measures are strongly
intercorrelated, with pairwise coefficients exceeding $0.8$, while the
dynamical indicators also exhibit substantial correlations with
structural position. In particular, Efficiency is strongly positively
correlated with degree and eigenvector centrality, with correlation
coefficients of approximately $0.86$ and $0.90$, respectively.
Distortion is strongly negatively correlated with closeness and
eigenvector centrality, with coefficients of approximately $-0.84$ and $-0.75$. Influence shows moderate to strong negative correlations with several structural measures, including approximately $-0.49$ with degree and $-0.71$ with closeness. 

The correlation structure indicates that, in the heterogeneous networks
considered here, node position strongly conditions the dynamical response. The correlations are substantial but not perfect, indicating that structural position does not uniquely determine the response-based indicators.

To assess the degree of the relationship between the dynamical and
structural descriptions, we perform a linear regression of each
dynamical indicator on the four topological measures. The quality of the
fit is quantified by the coefficient of determination, $R^2$, defined as
\begin{equation}
	R^2 =
	1-
	\frac{\sum_i (X_i-\hat{X}_i)^2}
	{\sum_i (X_i-\bar{X})^2},
\end{equation}
where $X_i$ represents the value of a dynamical indicator at node $i$,
$\hat{X}_i$ is the value predicted by the linear combination of the
structural measures, and $\bar{X}$ is the network mean. A larger
$R^2$ indicates that a larger fraction of the node-to-node variation in
the dynamical indicator is accounted for by the selected structural
measures.

The resulting values are $R^2(I)=0.581$, $R^2(E)=0.821$, and
$R^2(D)=0.782$ ($n=200$; see Table~\ref{tab:regression} in the
Appendix~\ref{sec:regression} for the full statistics). All three regressions are highly significant, with overall $p$ values below
$10^{-35}$. The structural measures therefore account for a substantial fraction of the variation in all three dynamical indicators, with the largest explanatory power found for Efficiency and the smallest for Influence. Equivalently, approximately $42\%$, $18\%$, and $22\%$ of the variance in $I$, $E$, and $D$, respectively, remains outside the linear prediction from the selected structural measures.

These results indicate that the dynamical and structural descriptions
are strongly coupled in the heterogeneous networks considered here.
The present framework therefore should not be interpreted as producing
node rankings that are statistically independent of topology.
Nevertheless, the construction provides a physically distinct
interpretation of node importance by separating the response into
distinct aspects.
The residual variance not explained by the selected linear structural model, together with the different physical meaning of the three indicators, provides a basis for distinguishing dynamical response characteristics even when their rankings are strongly influenced by structural position.

\subsection{Contribution matrix}

To resolve the aggregate response into source--target and temporal
components, we define a normalized contribution density for a given
source node $s$ as in Eq.~(\ref{eq:con}), and $I_s$ is defined as Eq.~(\ref{eq:influence}), so that the contribution density satisfies
\begin{equation}
	\sum_{i\neq s}
	\int_0^T C_{is}(t)\,dt
	=1.
\end{equation}
Thus, $C_{is}(t)$ represents the normalized contribution density
associated with target node $i$ at time $t$ for a perturbation applied
to source node $s$. The self-response $i=s$ is excluded consistently
with the definition of $I_s$.

\begin{figure*}[htbp]
	\centering
	\includegraphics[width=0.8\textwidth]{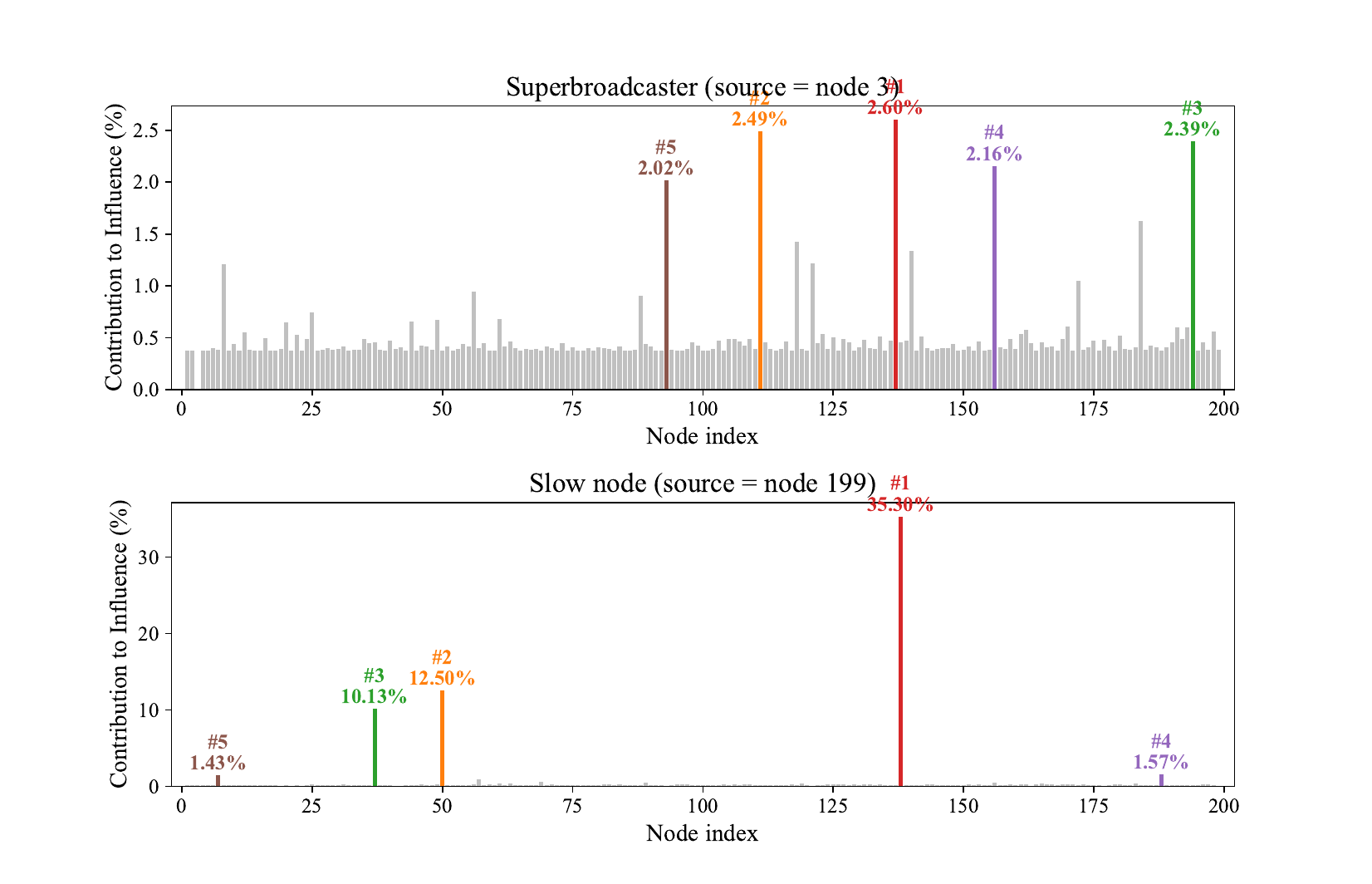}
	\caption{
		Time-integrated contribution profiles of a superbroadcaster and a slow
		node. The bars show the fractional contribution $P_{is}$ of each target
		node to the total Influence of the corresponding source node.
	}
	\label{fig:contribution}
\end{figure*}

Figure~\ref{fig:contribution} shows the time-integrated contribution
profile,
\begin{equation}
	P_{is}
	=
	\int_0^T C_{is}(t)\,dt,
\end{equation}
for a representative superbroadcaster and slow node. By construction,
$\sum_{i\neq s}P_{is}=1$, so $P_{is}$ gives the fractional contribution
of target node $i$ to the total Influence of source node $s$. The
comparison illustrates how the total response is distributed across
different target nodes for the two dynamical node classes.

\begin{figure*}[htbp]
	\centering
	\includegraphics[width=0.8\textwidth]{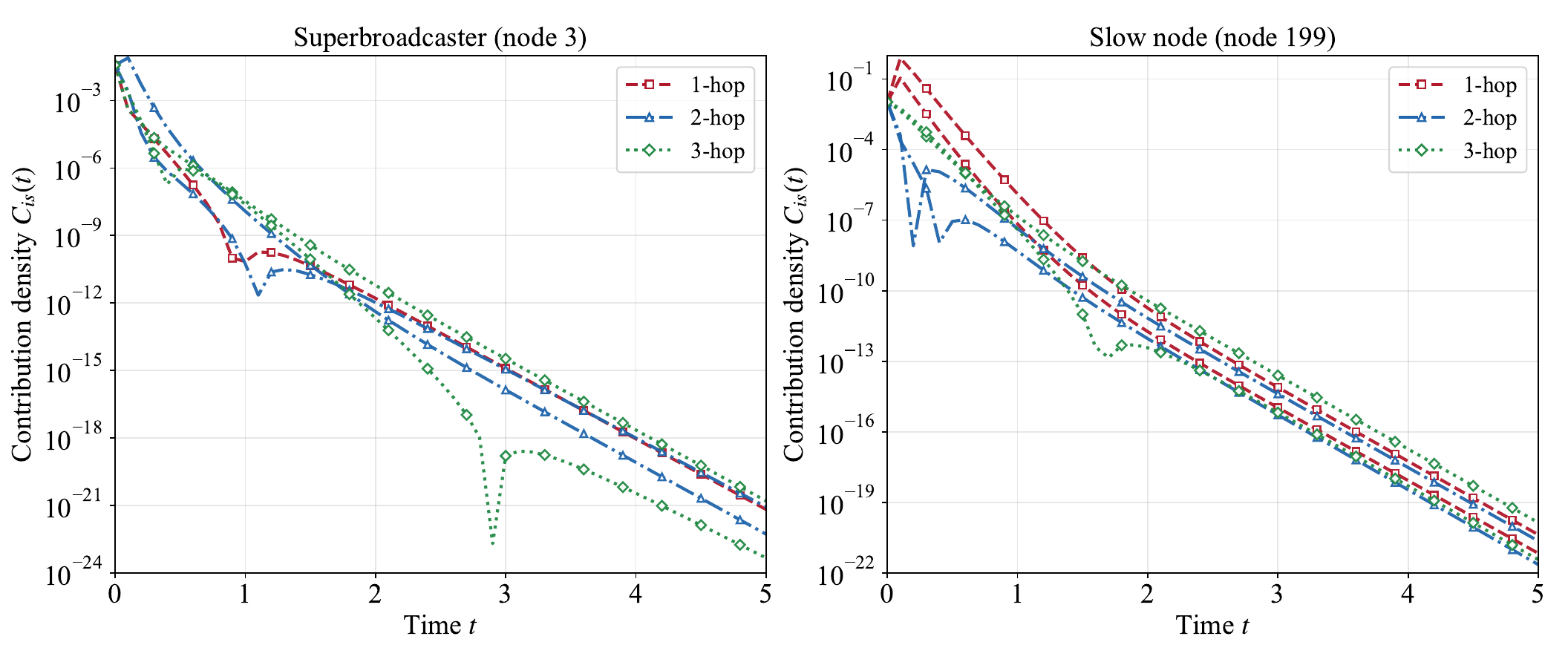}
	\caption{
		Early-time contribution density $C_{is}(t)$ for representative target
		nodes at one-, two-, and three-hop topological distances from the
		source. The source node itself is excluded. Curves are grouped by hop
		distance from the source.
	}
	\label{fig:trajectories}
\end{figure*}

The temporal structure of the same contribution density is illustrated
in Fig.~\ref{fig:trajectories}. For each source, representative target
nodes at one-, two-, and three-hop topological distances are selected,
and their contribution densities $C_{is}(t)$ are plotted over the
early-time interval. The source node itself is not included, consistent
with the definition of $C_{is}(t)$ for $i\neq s$. The resulting
trajectories resolve how the response develops at increasing
topological distances from the source. In the representative
superbroadcaster case, the response of targets at different hop
distances emerges over relatively similar time scales, whereas the slow
node exhibits more clearly separated temporal profiles across
successive hop distances.

The contribution representation therefore provides a direct way to
trace the aggregate Influence back to individual target nodes and
temporal intervals. The same source--target decomposition can be
extended to the temporal weighting used in Efficiency and to the
path-resolved spectral quantities used in Distortion; here we focus on
Influence as the representative example demonstrated numerically.

\subsection{Robustness across network realizations}

\begin{figure*}[htbp]
	\centering
	\includegraphics[width=0.9\textwidth]{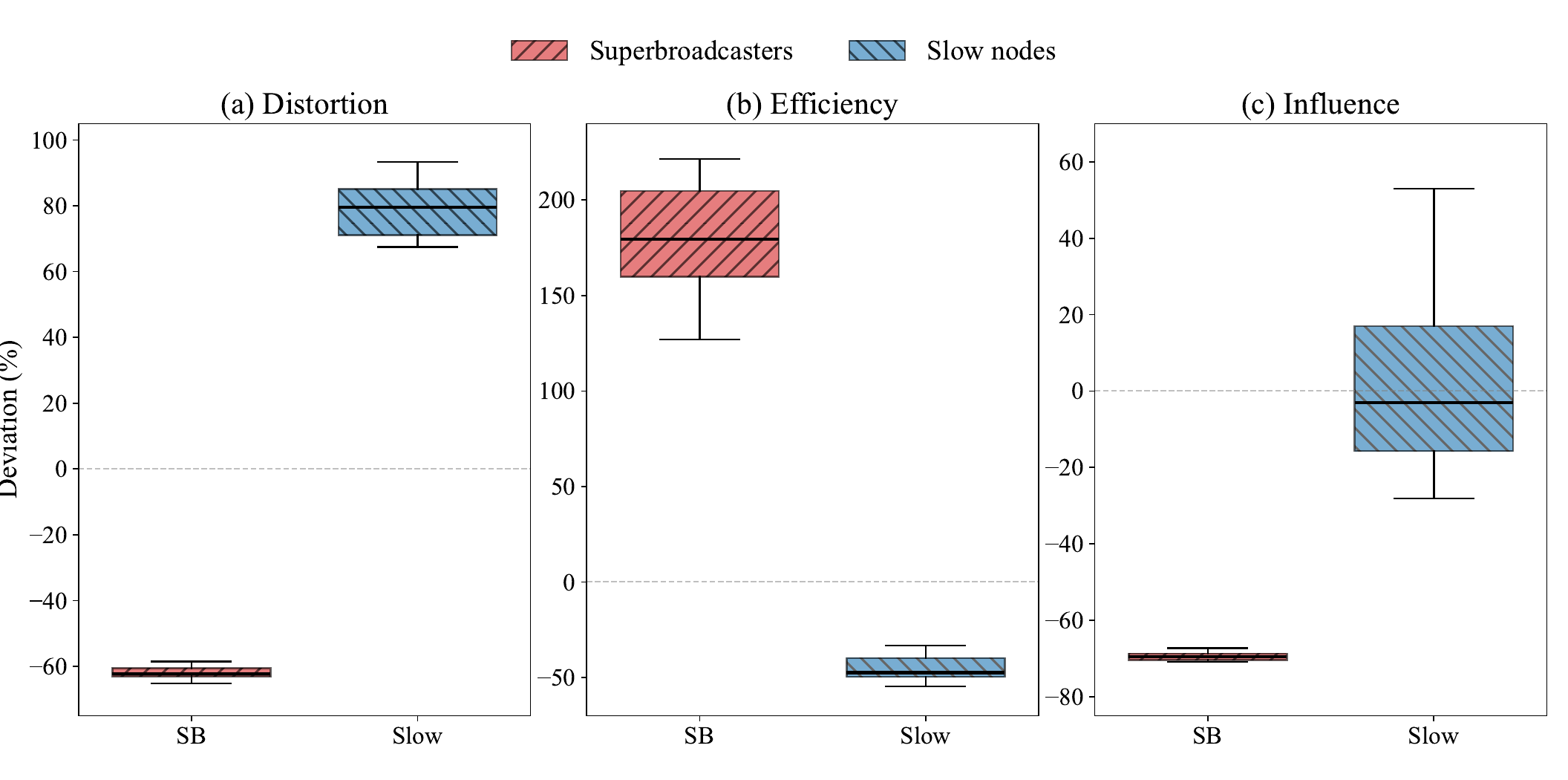}
	\caption{
		Robustness of the three dynamical indicators across $20$ independent BA
		network realizations. Boxes span the interquartile range; horizontal
		lines mark the median. Whiskers extend to the most extreme values within
		$1.5$ times the interquartile range.
	}
	\label{fig:robustness}
\end{figure*}

To assess the robustness of the three indicators with respect to network
realization, we repeated the simulation for 20 independent
Barab\'asi--Albert networks generated with different random seeds under
identical dynamical and numerical parameters. The resulting
distributions of the percentage deviations of the implanted
superbroadcasters and slow nodes from the mean of the unmodified nodes
are summarized in Fig.~\ref{fig:robustness}.

Distortion and Efficiency show complete separation between the two
implanted node classes across all 20 network realizations. For
superbroadcasters, the deviation in Distortion ranges approximately from $-65\%$ to $-57\%$, whereas the slow nodes exhibit deviations of about $+68\%$ to $+93\%$. Efficiency shows an equally clear separation: superbroadcasters exhibit positive deviations ranging from approximately $+127\%$ to $+222\%$, while slow nodes range from about $-55\%$ to $-33\%$. Influence displays a weaker but still systematic separation. The superbroadcasters remain consistently below the normal-node mean, with deviations of approximately $-71\%$ to $-66\%$, whereas the slow nodes show a broader range, from about $-28\%$ to $+53\%$. Thus, Efficiency and Distortion provide robust discrimination between the two implanted dynamical classes across network realizations, while Influence exhibits a stronger dependence on the particular network realization, especially for the slow-node class.

The contrasting behavior of the two node classes can be interpreted as the consequence of both structural embedding and local dynamical modification. Superbroadcasters are selected from the high-degree nodes and are simultaneously assigned enhanced coupling on selected edges and weaker damping. Their large positive Efficiency deviations therefore indicate substantially shorter characteristic response times, while their negative Influence deviations show that a faster response does not necessarily imply a larger cumulative response magnitude. Similarly, their strongly negative Distortion deviations indicate a broader spectral distribution of the propagated response according to the definition of $D$.

The slow nodes are selected from the lowest-degree nodes and are assigned both reduced coupling and stronger damping. Their peripheral structural position limits the number of available propagation pathways, while their local dynamical modification suppresses the response. Nevertheless, the resulting changes in Efficiency are relatively small, indicating that the characteristic response time is influenced by the broader network dynamics and not solely by the local damping parameter. The variation in Influence among slow nodes further demonstrates the sensitivity of the aggregate response to the specific topology of each
network realization.

Distortion exhibits a particularly clear class separation in the
robustness analysis. The lower $D$ values of the superbroadcasters
indicate a broader spectral distribution of their transmitted response,
whereas the larger $D$ values of the slow nodes correspond to stronger spectral concentration. This distinction is consistent across all 20 network realizations, while the magnitude of the difference should
be interpreted as a property of the present heterogeneous network
configuration rather than as a topology-independent signature.

\section{Discussion}

\subsection{Value of the multidimensional decomposition}

The motivation of the present framework is methodological rather than corrective. A dynamical network response is generally a structured object containing information about its magnitude, temporal evolution, and spectral organization. Reducing such a response to a single scalar necessarily compresses these different aspects into one quantity, making it difficult to determine which physical feature is responsible for a given change in node importance. The central idea of the present work is therefore to retain this internal structure by decomposing a single response into several physically interpretable components within a common mathematical construction. Such a decomposition does not seek
to replace established structural centralities, but to provide a way of
examining different physical aspects of dynamical node importance
without introducing separate models or unrelated measures for each
aspect.

An important consequence of this construction is that the resulting
description can remain closely connected to network topology without being reducible to a purely structural interpretation. The strong correlations observed in the present heterogeneous networks indicate that structural position is a major determinant of dynamical response, as expected for a networked system. At the same time, the response decomposition separates temporal and spectral organization that are not explicitly represented by a conventional topological coordinate. In this sense, the framework allows us to examine how the resulting perturbation response is organized in time and frequency, retaining physically meaningful distinctions.

The viewpoint also makes the framework modular for different
applications. Different components of the response can be examined
separately according to different fields, while their common origin ensures that they are comparable within the same dynamical system. The contribution matrix further extends this idea
from characterization to traceability by retaining information about
where and when the aggregate response is generated, and about its
spectral organization. This provides a direct route from a global node
descriptor back to the underlying source--target response structure,
which may be useful for network diagnosis, interpretation of
propagation mechanisms, and control theory.

\subsection{Scope of applicability}

Although the numerical demonstration focuses on Barab\'asi--Albert networks and the Kuramoto--Sakaguchi model, the indicators are defined from local node responses and do not depend on a particular global topology or on a specific choice of nonlinear dynamics. Other architectures---including Erd\H{o}s--R\'enyi, Watts--Strogatz, and modular networks---can be treated in the same way. In highly homogeneous networks, or in regimes where dynamical parameters vary only weakly, the indicators may exhibit little contrast between nodes; this is a natural consequence of low heterogeneity rather than a failure of the construction.

The formulation is likewise not restricted to oscillator dynamics. In principle, the same Green-function decomposition applies to any autonomous system that admits a stable reference state and a valid local linearization, including linear consensus dynamics~\cite{OlfatiSaber2007}, FitzHugh--Nagumo-type systems near a stable resting state~\cite{FitzHugh1961}, and epidemic models near a stable disease-free equilibrium~\cite{vandenDriessche2002}. These extensions require separate validation and are not demonstrated here.

The strong correlations observed between the dynamical and structural indicators further clarify the intended scope of the framework: it should not be read as a ranking independent of topology. Its contribution lies in the physical organization of the response into complementary dimensions, not in a claim of topological independence.

\subsection{Validity of the linear approximation}

As discussed in Sec.~II, the linear response framework retains only the first-order term in the expansion about the reference state. Its validity therefore requires that nonlinear contributions remain small compared with the linearized dynamics,
\begin{equation}
	|H[\delta\bm{\Gamma},\delta\bm{\Gamma}]| \ll |J\,\delta\bm{\Gamma}|.
\end{equation}
The range of validity depends on both the perturbation amplitude and the local structure of the reference state.

For a stable reference state, the non-neutral eigenvalues of $J$ must satisfy
\begin{equation}
	\max_{k\neq 0}\operatorname{Re}\lambda_k(J)<0,
\end{equation}
within numerical tolerance, with any neutral mode excluded when present. Even then, a non-normal Jacobian may produce transient amplification before asymptotic decay~\cite{Trefethen2005}. A practical diagnostic is
\begin{equation}
	G_{\max} = \sup_{t \geq 0} \|G(t)\|_2,
\end{equation}
where $\|\cdot\|_2$ is the induced Euclidean matrix norm. $G_{\max}$ measures the maximum amplification of an infinitesimal state perturbation under the linearized dynamics; it is used only as a check of applicability of the linear description, not as an additional centrality indicator.

Numerical divergences of the computed indicators can arise from two sources. Dynamical instability of the reference state,
\begin{equation}
	\max_{k\neq 0}\operatorname{Re}\lambda_k(J)>0,
\end{equation}
produces exponentially growing response components. Separately, even for a stable system, an excessively large time step relative to the fastest response timescale can destabilize the integration when heterogeneous couplings generate widely separated timescales~\cite{HairerWanner1996}, and direct evaluation of $\exp(Jt)$ can suffer overflow or loss of precision for long horizons or large $\|J\|$~\cite{Higham2008}. In practice, anomalous responses are diagnosed by verifying the spectrum of $J$ and by repeating the calculation with refined time steps.

\subsection{Limitations and future directions}

A central limitation is the requirement of a stable reference state at which the local linear response remains valid. Strong perturbations or operation far from that state can drive the response outside the linear neighborhood, rendering the Green-function description incomplete~\cite{Ruelle2009}. The present construction also assumes time-independent couplings and therefore does not address explicitly time-varying networks~\cite{Holme2012}.

These constraints suggest several extensions. First, the matrix Green function can be replaced by a time-ordered response kernel for general non-autonomous systems, or by a Floquet--Green function for periodically driven cases~\cite{floquet2005}. Second, higher-order response terms can be retained systematically to quantify the breakdown of linearity and to define effective nonlinear centralities~\cite{Ruelle2009}. Third, application to empirical data---power-grid snapshots, large-scale connectomes, or traffic networks---would test whether the additional dynamical dimensions improve the identification of functionally critical nodes beyond structural measures alone, and would help delimit the practical range of the linear-response assumption.

\subsection{Potential applications}

Because the indicators are defined from the response to a small perturbation about a nominal operating point, they are natural candidates for systems that are routinely monitored near a stable regime, without requiring a full nonlinear model beyond its local linearization.

In power grids, the indicators provide a dynamical complement to existing vulnerability analyses. Nodes with high Influence and low Distortion may correspond to locations whose perturbations propagate broadly while remaining comparatively spectrally dispersed, whereas high-Distortion nodes may preferentially emphasize a smaller set of oscillatory modes~\cite{coletta2016}. Such information is complementary to resistance-based or spectral rankings, which characterize vulnerability from structure but do not describe how a disturbance propagates through the dynamics.

In neuroscience, the same quantities could characterize how local perturbations shape resting-state activity or stimulus propagation~\cite{tang2018}: high-Influence, low-Distortion regions as efficient broadcasters, and high-Distortion regions as frequency-selective filters. Analogous interpretations apply to infrastructure and transportation networks under normal operation, where anomalous Efficiency or Distortion may flag latent bottlenecks.

These applications remain prospective. They require domain-specific validation and are not established by the present numerical study.

\section{Conclusion}

In this work, we establish a multidimensional methodology for analyzing
and tracing dynamical node importance from the Green function response of a linearized networked system. Rather than treating dynamical node importance as a single scalar quantity, we show how a common dynamical response can be systematically resolved into three complementary dimensions---Influence, Efficiency, and Distortion---and further analyzed through a contribution matrix that resolves source--target pathways and their temporal or spectral components. The resulting methodology provides a unified way to characterize how strongly, how rapidly, and with what degree of spectral concentration a perturbation propagates through a complex network.

The numerical results demonstrate that the three indicators distinguish
the implanted dynamical node classes through distinct physical
dimensions of the perturbation response. Efficiency and Distortion show complete separation between superbroadcasters and slow nodes across all 20 network realizations, while Influence provides a weaker but systematic separation and exhibits greater realization dependence,
particularly for the slow-node class. The correlation and regression
analyses further show that the three dynamical indicators remain
strongly coupled to conventional structural centralities. Although the framework does not possess statistical independence from topology, it provides a physically interpretable decomposition of node importance into cumulative response strength, temporal response rate, and spectral concentration. The contribution matrix further enables these aggregate descriptors to be traced back to specific target nodes, time intervals, and frequency components.

The formulation is not specific to a particular network topology or dynamical model, provided that a stable reference state exists and a local linear-response description is valid. The present numerical demonstration is restricted to weighted Kuramoto--Sakaguchi dynamics on Barab\'asi--Albert networks, and therefore does not by itself establish empirical generality across other systems. Strong perturbations, near-marginal stability, and time-dependent couplings remain outside the present construction and motivate extensions to nonlinear and nonstationary response regimes.

\section{Data Availability}
The data supporting the findings of this study are available from the corresponding author upon reasonable request. The source code used to perform the numerical simulations and generate the results is publicly available in Ref.~\cite{wang2026centrality}.

\appendix

\section{Regression Statistics}
\label{sec:regression}

Table~\ref{tab:regression} reports the full ordinary least-squares regression statistics underlying the $R^2$ values quoted in the main text. Each row corresponds to a regression of one dynamical measure on the four structural measures (degree, betweenness, closeness, and eigenvector centrality), with $n=200$ nodes and $k=4$ predictors. $R^2$ is the coefficient of determination, Adj.\ $R^2$ is the adjusted coefficient of determination, $F$ is the overall $F$ statistic, and $p$ is its associated $p$ value under the null hypothesis that all regression coefficients vanish.

\begin{table}[htbp]
	\centering
	\caption{Ordinary least-squares regressions of the dynamical measures on the four structural measures. All regressions use $n=200$ nodes and $k=4$ predictors. All three regressions are highly significant according to the overall $F$-tests, with $p<10^{-35}$}
	\label{tab:regression}
	\begin{ruledtabular}
		\begin{tabular}{lcccc}
			Dependent variable & $R^2$ & Adj.\ $R^2$ & $F$ & $p$ \\
			\hline
			$I$ & 0.581 & 0.572 & 67.5 & $9.3\times10^{-36}$ \\
			$E$ & 0.821 & 0.817 & 223.1 & $1.4\times10^{-71}$ \\
			$D$ & 0.782 & 0.777 & 174.7 & $2.7\times10^{-63}$ \\
		\end{tabular}
	\end{ruledtabular}
\end{table}

\section{Additional Numerical Parameters}
\label{sec:parameters}

The numerical parameters used in the simulations are
summarized here. All network realizations contain $N=200$ nodes and are generated using the Barab\'asi--Albert construction with attachment
parameter $m=4$. The global coupling strength is $K=3$, and the
Sakaguchi phase lag is $\alpha=0.1$. For ordinary nodes, the damping
parameter is fixed at $\gamma_i=0.1$. The three nodes with the highest
degree are designated as superbroadcasters and assigned
$\gamma_i=0.02$. For each superbroadcaster, five incident edges are assigned coupling weight $W_{ij}=2.0$. The three nodes with the lowest degree are designated as slow nodes and assigned $\gamma_i=0.5$, while all of their incident edges are assigned
$W_{ij}=0.5$. The remaining network edges have unit weight,
$W_{ij}=1$, and nonexistent edges have $W_{ij}=0$. The edge-weight
modifications are applied symmetrically to the two ends of each
undirected edge.

The Kuramoto dynamics are integrated with time step $\Delta t=0.01$
for at most $10\,000$ steps during the search for a phase-locked state.
For the subsequent Green function calculation, the response is evaluated over a time window $T_{\mathrm{green}}=50$ with sampling interval $\Delta t_{\mathrm{green}}=0.1$. Initial phases are independently sampled from a uniform distribution on $[0,2\pi)$, while the natural frequencies are drawn from $\mathcal{N}(0,1)$. The same numerical settings are used across network realizations, with the network seed varied between realizations.

\section{Finite-Time Truncation Check}
\label{sec:truncation}

The Green function integrals are evaluated over a finite time window. To verify that the chosen window is sufficiently long, we evaluate the response intensity
\begin{equation}
	R_s(t)=\sum_{i\neq s}|G_{is}(t)|^2
\end{equation}
for each source node $s$ and define the normalized residual response
\begin{equation}
	\rho_s(T)
	=
	\frac{R_s(T)}
	{\max_{0\leq t\leq T} R_s(t)}.
\end{equation}
We adopt the truncation criterion
\begin{equation}
	\max_s \rho_s(T)<10^{-3}.
\end{equation}
The numerical check is performed from the computed Green function time series using the same sampling interval $\Delta t_{\mathrm{green}}=0.1$ as in the main computation.

For the simulation reported in this paper, the Green function time series contains $501$ time samples over $0\leq t\leq50$. At $T=50$, the largest residual ratio among all $200$ source nodes is
\begin{equation}
	\max_s \rho_s(50)
	=
	3.54\times10^{-154},
\end{equation}
with a mean value of $3.17\times10^{-156}$ and a minimum value of
$1.31\times10^{-159}$. No source node exceeds the adopted threshold. The node with the largest residual ratio is
node $63$, for which
$\rho_{63}(50)=3.54\times10^{-154}$. The resulting check confirms that $T=50$ is sufficient for the numerical evaluation of the response integrals used in the study.

\nocite{*}
\bibliographystyle{apsrev4-2}
\bibliography{reference}

\begin{thebibliography}{52}%
\makeatletter
\providecommand \@ifxundefined [1]{%
 \@ifx{#1\undefined}
}%
\providecommand \@ifnum [1]{%
 \ifnum #1\expandafter \@firstoftwo
 \else \expandafter \@secondoftwo
 \fi
}%
\providecommand \@ifx [1]{%
 \ifx #1\expandafter \@firstoftwo
 \else \expandafter \@secondoftwo
 \fi
}%
\providecommand \natexlab [1]{#1}%
\providecommand \enquote  [1]{``#1''}%
\providecommand \bibnamefont  [1]{#1}%
\providecommand \bibfnamefont [1]{#1}%
\providecommand \citenamefont [1]{#1}%
\providecommand \href@noop [0]{\@secondoftwo}%
\providecommand \href [0]{\begingroup \@sanitize@url \@href}%
\providecommand \@href[1]{\@@startlink{#1}\@@href}%
\providecommand \@@href[1]{\endgroup#1\@@endlink}%
\providecommand \@sanitize@url [0]{\catcode `\\12\catcode `\$12\catcode
  `\&12\catcode `\#12\catcode `\^12\catcode `\_12\catcode `\%12\relax}%
\providecommand \@@startlink[1]{}%
\providecommand \@@endlink[0]{}%
\providecommand \url  [0]{\begingroup\@sanitize@url \@url }%
\providecommand \@url [1]{\endgroup\@href {#1}{\urlprefix }}%
\providecommand \urlprefix  [0]{URL }%
\providecommand \Eprint [0]{\href }%
\providecommand \doibase [0]{https://doi.org/}%
\providecommand \selectlanguage [0]{\@gobble}%
\providecommand \bibinfo  [0]{\@secondoftwo}%
\providecommand \bibfield  [0]{\@secondoftwo}%
\providecommand \translation [1]{[#1]}%
\providecommand \BibitemOpen [0]{}%
\providecommand \bibitemStop [0]{}%
\providecommand \bibitemNoStop [0]{.\EOS\space}%
\providecommand \EOS [0]{\spacefactor3000\relax}%
\providecommand \BibitemShut  [1]{\csname bibitem#1\endcsname}%
\let\auto@bib@innerbib\@empty
\bibitem [{\citenamefont {Albert}\ \emph {et~al.}(2000)\citenamefont {Albert},
  \citenamefont {Jeong},\ and\ \citenamefont {Barab{\'a}si}}]{albert2000}%
  \BibitemOpen
  \bibfield  {author} {\bibinfo {author} {\bibfnamefont {R.}~\bibnamefont
  {Albert}}, \bibinfo {author} {\bibfnamefont {H.}~\bibnamefont {Jeong}},\ and\
  \bibinfo {author} {\bibfnamefont {A.-L.}\ \bibnamefont {Barab{\'a}si}},\
  }\href {https://doi.org/10.1038/35019019} {\bibfield  {journal} {\bibinfo
  {journal} {Nature}\ }\textbf {\bibinfo {volume} {406}},\ \bibinfo {pages}
  {378} (\bibinfo {year} {2000})}\BibitemShut {NoStop}%
\bibitem [{\citenamefont {Pastor-Satorras}\ and\ \citenamefont
  {Vespignani}(2001)}]{pastor2001}%
  \BibitemOpen
  \bibfield  {author} {\bibinfo {author} {\bibfnamefont {R.}~\bibnamefont
  {Pastor-Satorras}}\ and\ \bibinfo {author} {\bibfnamefont {A.}~\bibnamefont
  {Vespignani}},\ }\href {https://doi.org/10.1103/PhysRevLett.86.3200}
  {\bibfield  {journal} {\bibinfo  {journal} {Phys. Rev. Lett.}\ }\textbf
  {\bibinfo {volume} {86}},\ \bibinfo {pages} {3200} (\bibinfo {year}
  {2001})}\BibitemShut {NoStop}%
\bibitem [{\citenamefont {Cohen}\ \emph {et~al.}(2001)\citenamefont {Cohen},
  \citenamefont {Erez}, \citenamefont {ben Avraham},\ and\ \citenamefont
  {Havlin}}]{cohen2001}%
  \BibitemOpen
  \bibfield  {author} {\bibinfo {author} {\bibfnamefont {R.}~\bibnamefont
  {Cohen}}, \bibinfo {author} {\bibfnamefont {K.}~\bibnamefont {Erez}},
  \bibinfo {author} {\bibfnamefont {D.}~\bibnamefont {ben Avraham}},\ and\
  \bibinfo {author} {\bibfnamefont {S.}~\bibnamefont {Havlin}},\ }\href
  {https://doi.org/10.1103/PhysRevLett.86.3682} {\bibfield  {journal} {\bibinfo
   {journal} {Phys. Rev. Lett.}\ }\textbf {\bibinfo {volume} {86}},\ \bibinfo
  {pages} {3682} (\bibinfo {year} {2001})}\BibitemShut {NoStop}%
\bibitem [{\citenamefont {Freeman}(1978)}]{freeman1978}%
  \BibitemOpen
  \bibfield  {author} {\bibinfo {author} {\bibfnamefont {L.~C.}\ \bibnamefont
  {Freeman}},\ }\href {https://doi.org/10.1016/0378-8733(78)90021-7} {\bibfield
   {journal} {\bibinfo  {journal} {Soc. Networks}\ }\textbf {\bibinfo {volume}
  {1}},\ \bibinfo {pages} {215} (\bibinfo {year} {1978})}\BibitemShut {NoStop}%
\bibitem [{\citenamefont {Bonacich}(1987)}]{bonacich1987}%
  \BibitemOpen
  \bibfield  {author} {\bibinfo {author} {\bibfnamefont {P.}~\bibnamefont
  {Bonacich}},\ }\href {https://doi.org/10.1086/228631} {\bibfield  {journal}
  {\bibinfo  {journal} {Am. J. Sociol.}\ }\textbf {\bibinfo {volume} {92}},\
  \bibinfo {pages} {1170} (\bibinfo {year} {1987})}\BibitemShut {NoStop}%
\bibitem [{\citenamefont {Newman}(2010)}]{newman2010}%
  \BibitemOpen
  \bibfield  {author} {\bibinfo {author} {\bibfnamefont {M.~E.~J.}\
  \bibnamefont {Newman}},\ }\href@noop {} {\emph {\bibinfo {title} {Networks:
  An Introduction}}}\ (\bibinfo  {publisher} {Oxford University Press},\
  \bibinfo {address} {Oxford},\ \bibinfo {year} {2010})\BibitemShut {NoStop}%
\bibitem [{\citenamefont {Oldham}\ \emph {et~al.}(2019)\citenamefont {Oldham},
  \citenamefont {Fulcher}, \citenamefont {Parkes}, \citenamefont
  {Arnatkeviciute}, \citenamefont {Suo},\ and\ \citenamefont
  {Fornito}}]{oldham2019}%
  \BibitemOpen
  \bibfield  {author} {\bibinfo {author} {\bibfnamefont {S.}~\bibnamefont
  {Oldham}}, \bibinfo {author} {\bibfnamefont {B.}~\bibnamefont {Fulcher}},
  \bibinfo {author} {\bibfnamefont {L.}~\bibnamefont {Parkes}}, \bibinfo
  {author} {\bibfnamefont {A.}~\bibnamefont {Arnatkeviciute}}, \bibinfo
  {author} {\bibfnamefont {C.}~\bibnamefont {Suo}},\ and\ \bibinfo {author}
  {\bibfnamefont {A.}~\bibnamefont {Fornito}},\ }\href
  {https://doi.org/10.1371/journal.pone.0220061} {\bibfield  {journal}
  {\bibinfo  {journal} {PLoS ONE}\ }\textbf {\bibinfo {volume} {14}},\ \bibinfo
  {pages} {e0220061} (\bibinfo {year} {2019})}\BibitemShut {NoStop}%
\bibitem [{\citenamefont {Klemm}\ \emph {et~al.}(2012)\citenamefont {Klemm},
  \citenamefont {Serrano}, \citenamefont {Egu{\'i}luz},\ and\ \citenamefont
  {San~Miguel}}]{Klemm2012}%
  \BibitemOpen
  \bibfield  {author} {\bibinfo {author} {\bibfnamefont {K.}~\bibnamefont
  {Klemm}}, \bibinfo {author} {\bibfnamefont {M.~{\'A}.}\ \bibnamefont
  {Serrano}}, \bibinfo {author} {\bibfnamefont {V.~M.}\ \bibnamefont
  {Egu{\'i}luz}},\ and\ \bibinfo {author} {\bibfnamefont {M.}~\bibnamefont
  {San~Miguel}},\ }\href {https://doi.org/10.1038/srep00292} {\bibfield
  {journal} {\bibinfo  {journal} {Sci. Rep.}\ }\textbf {\bibinfo {volume}
  {2}},\ \bibinfo {pages} {292} (\bibinfo {year} {2012})}\BibitemShut {NoStop}%
\bibitem [{\citenamefont {Harush}\ and\ \citenamefont
  {Barzel}(2017)}]{Harush2017}%
  \BibitemOpen
  \bibfield  {author} {\bibinfo {author} {\bibfnamefont {U.}~\bibnamefont
  {Harush}}\ and\ \bibinfo {author} {\bibfnamefont {B.}~\bibnamefont
  {Barzel}},\ }\href {https://doi.org/10.1038/s41467-017-01916-3} {\bibfield
  {journal} {\bibinfo  {journal} {Nat. Commun.}\ }\textbf {\bibinfo {volume}
  {8}},\ \bibinfo {pages} {2181} (\bibinfo {year} {2017})}\BibitemShut
  {NoStop}%
\bibitem [{\citenamefont {van Elteren}\ \emph {et~al.}(2022)\citenamefont {van
  Elteren}, \citenamefont {Quax},\ and\ \citenamefont {Sloot}}]{Elteren2022}%
  \BibitemOpen
  \bibfield  {author} {\bibinfo {author} {\bibfnamefont {C.}~\bibnamefont {van
  Elteren}}, \bibinfo {author} {\bibfnamefont {R.}~\bibnamefont {Quax}},\ and\
  \bibinfo {author} {\bibfnamefont {P.}~\bibnamefont {Sloot}},\ }\href
  {https://doi.org/10.1016/j.physa.2022.126889} {\bibfield  {journal} {\bibinfo
   {journal} {Physica A}\ }\textbf {\bibinfo {volume} {593}},\ \bibinfo {pages}
  {126889} (\bibinfo {year} {2022})}\BibitemShut {NoStop}%
\bibitem [{\citenamefont {Restrepo}\ \emph {et~al.}(2006)\citenamefont
  {Restrepo}, \citenamefont {Ott},\ and\ \citenamefont {Hunt}}]{restrepo2006}%
  \BibitemOpen
  \bibfield  {author} {\bibinfo {author} {\bibfnamefont {J.~G.}\ \bibnamefont
  {Restrepo}}, \bibinfo {author} {\bibfnamefont {E.}~\bibnamefont {Ott}},\ and\
  \bibinfo {author} {\bibfnamefont {B.~R.}\ \bibnamefont {Hunt}},\ }\href
  {https://doi.org/10.1103/PhysRevLett.97.094102} {\bibfield  {journal}
  {\bibinfo  {journal} {Phys. Rev. Lett.}\ }\textbf {\bibinfo {volume} {97}},\
  \bibinfo {pages} {094102} (\bibinfo {year} {2006})}\BibitemShut {NoStop}%
\bibitem [{\citenamefont {Liu}\ \emph {et~al.}(2011)\citenamefont {Liu},
  \citenamefont {Slotine},\ and\ \citenamefont {Barab{\'a}si}}]{liu2011}%
  \BibitemOpen
  \bibfield  {author} {\bibinfo {author} {\bibfnamefont {Y.-Y.}\ \bibnamefont
  {Liu}}, \bibinfo {author} {\bibfnamefont {J.-J.}\ \bibnamefont {Slotine}},\
  and\ \bibinfo {author} {\bibfnamefont {A.-L.}\ \bibnamefont {Barab{\'a}si}},\
  }\href {https://doi.org/10.1038/nature10011} {\bibfield  {journal} {\bibinfo
  {journal} {Nature}\ }\textbf {\bibinfo {volume} {473}},\ \bibinfo {pages}
  {167} (\bibinfo {year} {2011})}\BibitemShut {NoStop}%
\bibitem [{\citenamefont {Motter}\ and\ \citenamefont
  {Lai}(2002)}]{motter2002}%
  \BibitemOpen
  \bibfield  {author} {\bibinfo {author} {\bibfnamefont {A.~E.}\ \bibnamefont
  {Motter}}\ and\ \bibinfo {author} {\bibfnamefont {Y.-C.}\ \bibnamefont
  {Lai}},\ }\href {https://doi.org/10.1103/PhysRevE.66.065102} {\bibfield
  {journal} {\bibinfo  {journal} {Phys. Rev. E}\ }\textbf {\bibinfo {volume}
  {66}},\ \bibinfo {pages} {065102} (\bibinfo {year} {2002})}\BibitemShut
  {NoStop}%
\bibitem [{\citenamefont {Kitsak}\ \emph {et~al.}(2010)\citenamefont {Kitsak},
  \citenamefont {Gallos}, \citenamefont {Havlin}, \citenamefont {Liljeros},
  \citenamefont {Muchnik}, \citenamefont {Stanley},\ and\ \citenamefont
  {Makse}}]{kitsak2010}%
  \BibitemOpen
  \bibfield  {author} {\bibinfo {author} {\bibfnamefont {M.}~\bibnamefont
  {Kitsak}}, \bibinfo {author} {\bibfnamefont {L.~K.}\ \bibnamefont {Gallos}},
  \bibinfo {author} {\bibfnamefont {S.}~\bibnamefont {Havlin}}, \bibinfo
  {author} {\bibfnamefont {F.}~\bibnamefont {Liljeros}}, \bibinfo {author}
  {\bibfnamefont {L.}~\bibnamefont {Muchnik}}, \bibinfo {author} {\bibfnamefont
  {H.~E.}\ \bibnamefont {Stanley}},\ and\ \bibinfo {author} {\bibfnamefont
  {H.~A.}\ \bibnamefont {Makse}},\ }\href {https://doi.org/10.1038/nphys1746}
  {\bibfield  {journal} {\bibinfo  {journal} {Nat. Phys.}\ }\textbf {\bibinfo
  {volume} {6}},\ \bibinfo {pages} {888} (\bibinfo {year} {2010})}\BibitemShut
  {NoStop}%
\bibitem [{\citenamefont {Estrada}\ and\ \citenamefont
  {Rodriguez-Velazquez}(2005)}]{estrada2005}%
  \BibitemOpen
  \bibfield  {author} {\bibinfo {author} {\bibfnamefont {E.}~\bibnamefont
  {Estrada}}\ and\ \bibinfo {author} {\bibfnamefont {J.~A.}\ \bibnamefont
  {Rodriguez-Velazquez}},\ }\href {https://doi.org/10.1103/PhysRevE.71.056103}
  {\bibfield  {journal} {\bibinfo  {journal} {Phys. Rev. E}\ }\textbf {\bibinfo
  {volume} {71}},\ \bibinfo {pages} {056103} (\bibinfo {year}
  {2005})}\BibitemShut {NoStop}%
\bibitem [{\citenamefont {Estrada}\ and\ \citenamefont
  {Hatano}(2008)}]{estrada2008}%
  \BibitemOpen
  \bibfield  {author} {\bibinfo {author} {\bibfnamefont {E.}~\bibnamefont
  {Estrada}}\ and\ \bibinfo {author} {\bibfnamefont {N.}~\bibnamefont
  {Hatano}},\ }\href {https://doi.org/10.1103/PhysRevE.77.036111} {\bibfield
  {journal} {\bibinfo  {journal} {Phys. Rev. E}\ }\textbf {\bibinfo {volume}
  {77}},\ \bibinfo {pages} {036111} (\bibinfo {year} {2008})}\BibitemShut
  {NoStop}%
\bibitem [{\citenamefont {Gilson}\ \emph
  {et~al.}(2018{\natexlab{a}})\citenamefont {Gilson}, \citenamefont {Kouvaris},
  \citenamefont {Deco},\ and\ \citenamefont
  {Zamora-L{\'o}pez}}]{gilson2018framework}%
  \BibitemOpen
  \bibfield  {author} {\bibinfo {author} {\bibfnamefont {M.}~\bibnamefont
  {Gilson}}, \bibinfo {author} {\bibfnamefont {N.~E.}\ \bibnamefont
  {Kouvaris}}, \bibinfo {author} {\bibfnamefont {G.}~\bibnamefont {Deco}},\
  and\ \bibinfo {author} {\bibfnamefont {G.}~\bibnamefont {Zamora-L{\'o}pez}},\
  }\href {https://doi.org/10.1103/PhysRevE.97.052301} {\bibfield  {journal}
  {\bibinfo  {journal} {Phys. Rev. E}\ }\textbf {\bibinfo {volume} {97}},\
  \bibinfo {pages} {052301} (\bibinfo {year} {2018}{\natexlab{a}})}\BibitemShut
  {NoStop}%
\bibitem [{\citenamefont {Gilson}\ \emph
  {et~al.}(2018{\natexlab{b}})\citenamefont {Gilson}, \citenamefont {Deco},
  \citenamefont {Friston}, \citenamefont {Hagmann}, \citenamefont {Mantini},
  \citenamefont {Betti}, \citenamefont {Romani},\ and\ \citenamefont
  {Corbetta}}]{gilson2018neuro}%
  \BibitemOpen
  \bibfield  {author} {\bibinfo {author} {\bibfnamefont {M.}~\bibnamefont
  {Gilson}}, \bibinfo {author} {\bibfnamefont {G.}~\bibnamefont {Deco}},
  \bibinfo {author} {\bibfnamefont {K.~J.}\ \bibnamefont {Friston}}, \bibinfo
  {author} {\bibfnamefont {P.}~\bibnamefont {Hagmann}}, \bibinfo {author}
  {\bibfnamefont {D.}~\bibnamefont {Mantini}}, \bibinfo {author} {\bibfnamefont
  {V.}~\bibnamefont {Betti}}, \bibinfo {author} {\bibfnamefont {G.~L.}\
  \bibnamefont {Romani}},\ and\ \bibinfo {author} {\bibfnamefont
  {M.}~\bibnamefont {Corbetta}},\ }\href
  {https://doi.org/10.1016/j.neuroimage.2017.09.061} {\bibfield  {journal}
  {\bibinfo  {journal} {NeuroImage}\ }\textbf {\bibinfo {volume} {180}},\
  \bibinfo {pages} {534} (\bibinfo {year} {2018}{\natexlab{b}})}\BibitemShut
  {NoStop}%
\bibitem [{\citenamefont {Bartesaghi}\ \emph {et~al.}(2022)\citenamefont
  {Bartesaghi}, \citenamefont {Clemente},\ and\ \citenamefont
  {Grassi}}]{bartesaghi2022}%
  \BibitemOpen
  \bibfield  {author} {\bibinfo {author} {\bibfnamefont {P.}~\bibnamefont
  {Bartesaghi}}, \bibinfo {author} {\bibfnamefont {G.~P.}\ \bibnamefont
  {Clemente}},\ and\ \bibinfo {author} {\bibfnamefont {R.}~\bibnamefont
  {Grassi}},\ }\href {https://doi.org/10.1007/s11403-020-00309-y} {\bibfield
  {journal} {\bibinfo  {journal} {J. Econ. Interact. Coord.}\ }\textbf
  {\bibinfo {volume} {17}},\ \bibinfo {pages} {405} (\bibinfo {year}
  {2022})}\BibitemShut {NoStop}%
\bibitem [{\citenamefont {Kubo}(1957)}]{Kubo1957}%
  \BibitemOpen
  \bibfield  {author} {\bibinfo {author} {\bibfnamefont {R.}~\bibnamefont
  {Kubo}},\ }\href {https://doi.org/10.1143/JPSJ.12.570} {\bibfield  {journal}
  {\bibinfo  {journal} {J. Phys. Soc. Jpn.}\ }\textbf {\bibinfo {volume}
  {12}},\ \bibinfo {pages} {570} (\bibinfo {year} {1957})}\BibitemShut
  {NoStop}%
\bibitem [{\citenamefont {Kubo}(1966)}]{Kubo1966}%
  \BibitemOpen
  \bibfield  {author} {\bibinfo {author} {\bibfnamefont {R.}~\bibnamefont
  {Kubo}},\ }\href {https://doi.org/10.1088/0034-4885/29/1/306} {\bibfield
  {journal} {\bibinfo  {journal} {Rep. Prog. Phys.}\ }\textbf {\bibinfo
  {volume} {29}},\ \bibinfo {pages} {255} (\bibinfo {year} {1966})}\BibitemShut
  {NoStop}%
\bibitem [{\citenamefont {Ruelle}(1998)}]{Ruelle1998}%
  \BibitemOpen
  \bibfield  {author} {\bibinfo {author} {\bibfnamefont {D.}~\bibnamefont
  {Ruelle}},\ }\href {https://doi.org/10.1016/S0375-9601(98)00419-8} {\bibfield
   {journal} {\bibinfo  {journal} {Phys. Lett. A}\ }\textbf {\bibinfo {volume}
  {245}},\ \bibinfo {pages} {220} (\bibinfo {year} {1998})}\BibitemShut
  {NoStop}%
\bibitem [{\citenamefont {Ruelle}(2009)}]{Ruelle2009}%
  \BibitemOpen
  \bibfield  {author} {\bibinfo {author} {\bibfnamefont {D.}~\bibnamefont
  {Ruelle}},\ }\href {https://doi.org/10.1088/0951-7715/22/4/009} {\bibfield
  {journal} {\bibinfo  {journal} {Nonlinearity}\ }\textbf {\bibinfo {volume}
  {22}},\ \bibinfo {pages} {855} (\bibinfo {year} {2009})}\BibitemShut
  {NoStop}%
\bibitem [{\citenamefont {Thouless}(1974)}]{Thouless1974}%
  \BibitemOpen
  \bibfield  {author} {\bibinfo {author} {\bibfnamefont {D.~J.}\ \bibnamefont
  {Thouless}},\ }\href {https://doi.org/10.1016/0370-1573(74)90029-5}
  {\bibfield  {journal} {\bibinfo  {journal} {Phys. Rep.}\ }\textbf {\bibinfo
  {volume} {13}},\ \bibinfo {pages} {93} (\bibinfo {year} {1974})}\BibitemShut
  {NoStop}%
\bibitem [{\citenamefont {Wegner}(1980)}]{Wegner1980}%
  \BibitemOpen
  \bibfield  {author} {\bibinfo {author} {\bibfnamefont {F.}~\bibnamefont
  {Wegner}},\ }\href {https://doi.org/10.1007/BF01325284} {\bibfield  {journal}
  {\bibinfo  {journal} {Z. Phys. B}\ }\textbf {\bibinfo {volume} {36}},\
  \bibinfo {pages} {209} (\bibinfo {year} {1980})}\BibitemShut {NoStop}%
\bibitem [{\citenamefont {Bell}\ and\ \citenamefont
  {Dean}(1970)}]{BellDean1970}%
  \BibitemOpen
  \bibfield  {author} {\bibinfo {author} {\bibfnamefont {R.~J.}\ \bibnamefont
  {Bell}}\ and\ \bibinfo {author} {\bibfnamefont {P.}~\bibnamefont {Dean}},\
  }\href {https://doi.org/10.1039/DF9705000055} {\bibfield  {journal} {\bibinfo
   {journal} {Discuss. Faraday Soc.}\ }\textbf {\bibinfo {volume} {50}},\
  \bibinfo {pages} {55} (\bibinfo {year} {1970})}\BibitemShut {NoStop}%
\bibitem [{\citenamefont {Bendat}\ and\ \citenamefont
  {Piersol}(2010)}]{bendat2010}%
  \BibitemOpen
  \bibfield  {author} {\bibinfo {author} {\bibfnamefont {J.~S.}\ \bibnamefont
  {Bendat}}\ and\ \bibinfo {author} {\bibfnamefont {A.~G.}\ \bibnamefont
  {Piersol}},\ }\href@noop {} {\emph {\bibinfo {title} {Random Data: Analysis
  and Measurement Procedures}}},\ \bibinfo {edition} {4th}\ ed.\ (\bibinfo
  {publisher} {Wiley},\ \bibinfo {address} {Hoboken, NJ},\ \bibinfo {year}
  {2010})\BibitemShut {NoStop}%
\bibitem [{\citenamefont {Percival}\ and\ \citenamefont
  {Walden}(1993)}]{percival1993}%
  \BibitemOpen
  \bibfield  {author} {\bibinfo {author} {\bibfnamefont {D.~B.}\ \bibnamefont
  {Percival}}\ and\ \bibinfo {author} {\bibfnamefont {A.~T.}\ \bibnamefont
  {Walden}},\ }\href@noop {} {\emph {\bibinfo {title} {Spectral Analysis for
  Physical Applications}}}\ (\bibinfo  {publisher} {Cambridge University
  Press},\ \bibinfo {address} {Cambridge},\ \bibinfo {year} {1993})\BibitemShut
  {NoStop}%
\bibitem [{\citenamefont {Acebr{\'o}n}\ \emph {et~al.}(2005)\citenamefont
  {Acebr{\'o}n}, \citenamefont {Bonilla}, \citenamefont {P{\'e}rez~Vicente},
  \citenamefont {Ritort},\ and\ \citenamefont {Spigler}}]{acebron2005}%
  \BibitemOpen
  \bibfield  {author} {\bibinfo {author} {\bibfnamefont {J.~A.}\ \bibnamefont
  {Acebr{\'o}n}}, \bibinfo {author} {\bibfnamefont {L.~L.}\ \bibnamefont
  {Bonilla}}, \bibinfo {author} {\bibfnamefont {C.~J.}\ \bibnamefont
  {P{\'e}rez~Vicente}}, \bibinfo {author} {\bibfnamefont {F.}~\bibnamefont
  {Ritort}},\ and\ \bibinfo {author} {\bibfnamefont {R.}~\bibnamefont
  {Spigler}},\ }\href {https://doi.org/10.1103/RevModPhys.77.137} {\bibfield
  {journal} {\bibinfo  {journal} {Rev. Mod. Phys.}\ }\textbf {\bibinfo {volume}
  {77}},\ \bibinfo {pages} {137} (\bibinfo {year} {2005})}\BibitemShut
  {NoStop}%
\bibitem [{\citenamefont {Barab{\'a}si}\ and\ \citenamefont
  {Albert}(1999)}]{barabasi1999}%
  \BibitemOpen
  \bibfield  {author} {\bibinfo {author} {\bibfnamefont {A.-L.}\ \bibnamefont
  {Barab{\'a}si}}\ and\ \bibinfo {author} {\bibfnamefont {R.}~\bibnamefont
  {Albert}},\ }\href {https://doi.org/10.1126/science.286.5439.509} {\bibfield
  {journal} {\bibinfo  {journal} {Science}\ }\textbf {\bibinfo {volume}
  {286}},\ \bibinfo {pages} {509} (\bibinfo {year} {1999})}\BibitemShut
  {NoStop}%
\bibitem [{\citenamefont {Craddock}\ \emph {et~al.}(2012)\citenamefont
  {Craddock}, \citenamefont {James}, \citenamefont {Holtzheimer}, \citenamefont
  {Hu},\ and\ \citenamefont {Mayberg}}]{Craddock2012}%
  \BibitemOpen
  \bibfield  {author} {\bibinfo {author} {\bibfnamefont {R.~C.}\ \bibnamefont
  {Craddock}}, \bibinfo {author} {\bibfnamefont {G.~A.}\ \bibnamefont {James}},
  \bibinfo {author} {\bibfnamefont {P.~E.}\ \bibnamefont {Holtzheimer}},
  \bibinfo {author} {\bibfnamefont {X.~P.}\ \bibnamefont {Hu}},\ and\ \bibinfo
  {author} {\bibfnamefont {H.~S.}\ \bibnamefont {Mayberg}},\ }\href
  {https://doi.org/10.1002/hbm.21333} {\bibfield  {journal} {\bibinfo
  {journal} {Hum. Brain Mapp.}\ }\textbf {\bibinfo {volume} {33}},\ \bibinfo
  {pages} {1914} (\bibinfo {year} {2012})}\BibitemShut {NoStop}%
\bibitem [{\citenamefont {Schaefer}\ \emph {et~al.}(2018)\citenamefont
  {Schaefer}, \citenamefont {Kong}, \citenamefont {Gordon}, \citenamefont
  {Laumann}, \citenamefont {Zuo}, \citenamefont {Holmes}, \citenamefont
  {Eickhoff},\ and\ \citenamefont {Yeo}}]{Schaefer2018}%
  \BibitemOpen
  \bibfield  {author} {\bibinfo {author} {\bibfnamefont {A.}~\bibnamefont
  {Schaefer}}, \bibinfo {author} {\bibfnamefont {R.}~\bibnamefont {Kong}},
  \bibinfo {author} {\bibfnamefont {E.~M.}\ \bibnamefont {Gordon}}, \bibinfo
  {author} {\bibfnamefont {T.~O.}\ \bibnamefont {Laumann}}, \bibinfo {author}
  {\bibfnamefont {X.-N.}\ \bibnamefont {Zuo}}, \bibinfo {author} {\bibfnamefont
  {A.~J.}\ \bibnamefont {Holmes}}, \bibinfo {author} {\bibfnamefont {S.~B.}\
  \bibnamefont {Eickhoff}},\ and\ \bibinfo {author} {\bibfnamefont {B.~T.~T.}\
  \bibnamefont {Yeo}},\ }\href {https://doi.org/10.1093/cercor/bhx179}
  {\bibfield  {journal} {\bibinfo  {journal} {Cereb. Cortex}\ }\textbf
  {\bibinfo {volume} {28}},\ \bibinfo {pages} {3095} (\bibinfo {year}
  {2018})}\BibitemShut {NoStop}%
\bibitem [{\citenamefont {Clemente}\ \emph {et~al.}(2023)\citenamefont
  {Clemente}, \citenamefont {Cornaro},\ and\ \citenamefont
  {Della~Corte}}]{clemente2023}%
  \BibitemOpen
  \bibfield  {author} {\bibinfo {author} {\bibfnamefont {G.~P.}\ \bibnamefont
  {Clemente}}, \bibinfo {author} {\bibfnamefont {A.}~\bibnamefont {Cornaro}},\
  and\ \bibinfo {author} {\bibfnamefont {F.}~\bibnamefont {Della~Corte}},\
  }\href {https://doi.org/10.1038/s41598-023-41038-z} {\bibfield  {journal}
  {\bibinfo  {journal} {Sci. Rep.}\ }\textbf {\bibinfo {volume} {13}},\
  \bibinfo {pages} {13966} (\bibinfo {year} {2023})}\BibitemShut {NoStop}%
\bibitem [{\citenamefont {Pagani}\ and\ \citenamefont
  {Aiello}(2014)}]{Pagani2014}%
  \BibitemOpen
  \bibfield  {author} {\bibinfo {author} {\bibfnamefont {G.~A.}\ \bibnamefont
  {Pagani}}\ and\ \bibinfo {author} {\bibfnamefont {M.}~\bibnamefont
  {Aiello}},\ }\href {https://doi.org/10.1016/j.physa.2013.11.022} {\bibfield
  {journal} {\bibinfo  {journal} {Physica A}\ }\textbf {\bibinfo {volume}
  {396}},\ \bibinfo {pages} {248} (\bibinfo {year} {2014})}\BibitemShut
  {NoStop}%
\bibitem [{\citenamefont {Bonett}\ and\ \citenamefont
  {Wright}(2000)}]{Bonett2000}%
  \BibitemOpen
  \bibfield  {author} {\bibinfo {author} {\bibfnamefont {D.~G.}\ \bibnamefont
  {Bonett}}\ and\ \bibinfo {author} {\bibfnamefont {T.~A.}\ \bibnamefont
  {Wright}},\ }\href {https://doi.org/10.1007/BF02294183} {\bibfield  {journal}
  {\bibinfo  {journal} {Psychometrika}\ }\textbf {\bibinfo {volume} {65}},\
  \bibinfo {pages} {23} (\bibinfo {year} {2000})}\BibitemShut {NoStop}%
\bibitem [{\citenamefont {Higham}(2008)}]{Higham2008}%
  \BibitemOpen
  \bibfield  {author} {\bibinfo {author} {\bibfnamefont {N.~J.}\ \bibnamefont
  {Higham}},\ }\href {https://doi.org/10.1137/1.9780898717778} {\emph {\bibinfo
  {title} {Functions of Matrices: Theory and Computation}}}\ (\bibinfo
  {publisher} {SIAM},\ \bibinfo {year} {2008})\BibitemShut {NoStop}%
\bibitem [{\citenamefont {Al-Mohy}\ and\ \citenamefont
  {Higham}(2011)}]{AlMohy2011}%
  \BibitemOpen
  \bibfield  {author} {\bibinfo {author} {\bibfnamefont {A.~H.}\ \bibnamefont
  {Al-Mohy}}\ and\ \bibinfo {author} {\bibfnamefont {N.~J.}\ \bibnamefont
  {Higham}},\ }\href {https://doi.org/10.1137/100788860} {\bibfield  {journal}
  {\bibinfo  {journal} {SIAM J. Sci. Comput.}\ }\textbf {\bibinfo {volume}
  {33}},\ \bibinfo {pages} {488} (\bibinfo {year} {2011})}\BibitemShut
  {NoStop}%
\bibitem [{\citenamefont {Benzi}\ and\ \citenamefont
  {Boito}(2020)}]{BenziBoito2020}%
  \BibitemOpen
  \bibfield  {author} {\bibinfo {author} {\bibfnamefont {M.}~\bibnamefont
  {Benzi}}\ and\ \bibinfo {author} {\bibfnamefont {P.}~\bibnamefont {Boito}},\
  }\href {https://doi.org/10.1002/gamm.202000012} {\bibfield  {journal}
  {\bibinfo  {journal} {GAMM-Mitteilungen}\ }\textbf {\bibinfo {volume} {43}},\
  \bibinfo {pages} {e202000012} (\bibinfo {year} {2020})}\BibitemShut {NoStop}%
\bibitem [{\citenamefont {Moreira}\ and\ \citenamefont
  {de~Aguiar}(2019)}]{Moreira2019}%
  \BibitemOpen
  \bibfield  {author} {\bibinfo {author} {\bibfnamefont {C.~A.}\ \bibnamefont
  {Moreira}}\ and\ \bibinfo {author} {\bibfnamefont {M.~A.~M.}\ \bibnamefont
  {de~Aguiar}},\ }\href {https://doi.org/10.1016/j.physa.2018.09.096}
  {\bibfield  {journal} {\bibinfo  {journal} {Physica A}\ }\textbf {\bibinfo
  {volume} {514}},\ \bibinfo {pages} {487} (\bibinfo {year}
  {2019})}\BibitemShut {NoStop}%
\bibitem [{\citenamefont {Novelli}\ and\ \citenamefont
  {Lizier}(2021)}]{Novelli2021}%
  \BibitemOpen
  \bibfield  {author} {\bibinfo {author} {\bibfnamefont {L.}~\bibnamefont
  {Novelli}}\ and\ \bibinfo {author} {\bibfnamefont {J.~T.}\ \bibnamefont
  {Lizier}},\ }\href {https://doi.org/10.1162/netn_a_00178} {\bibfield
  {journal} {\bibinfo  {journal} {Netw. Neurosci.}\ }\textbf {\bibinfo {volume}
  {5}},\ \bibinfo {pages} {373} (\bibinfo {year} {2021})}\BibitemShut {NoStop}%
\bibitem [{\citenamefont {Sakaguchi}\ and\ \citenamefont
  {Kuramoto}(1986)}]{sakaguchi1986}%
  \BibitemOpen
  \bibfield  {author} {\bibinfo {author} {\bibfnamefont {H.}~\bibnamefont
  {Sakaguchi}}\ and\ \bibinfo {author} {\bibfnamefont {Y.}~\bibnamefont
  {Kuramoto}},\ }\href {https://doi.org/10.1143/PTP.76.576} {\bibfield
  {journal} {\bibinfo  {journal} {Prog. Theor. Phys.}\ }\textbf {\bibinfo
  {volume} {76}},\ \bibinfo {pages} {576} (\bibinfo {year} {1986})}\BibitemShut
  {NoStop}%
\bibitem [{\citenamefont {Zhou}\ \emph {et~al.}(2006)\citenamefont {Zhou},
  \citenamefont {Motter},\ and\ \citenamefont {Kurths}}]{zhou2006}%
  \BibitemOpen
  \bibfield  {author} {\bibinfo {author} {\bibfnamefont {C.}~\bibnamefont
  {Zhou}}, \bibinfo {author} {\bibfnamefont {A.~E.}\ \bibnamefont {Motter}},\
  and\ \bibinfo {author} {\bibfnamefont {J.}~\bibnamefont {Kurths}},\ }\href
  {https://doi.org/10.1103/PhysRevLett.96.034101} {\bibfield  {journal}
  {\bibinfo  {journal} {Phys. Rev. Lett.}\ }\textbf {\bibinfo {volume} {96}},\
  \bibinfo {pages} {034101} (\bibinfo {year} {2006})}\BibitemShut {NoStop}%
\bibitem [{\citenamefont {Olfati-Saber}\ \emph {et~al.}(2007)\citenamefont
  {Olfati-Saber}, \citenamefont {Fax},\ and\ \citenamefont
  {Murray}}]{OlfatiSaber2007}%
  \BibitemOpen
  \bibfield  {author} {\bibinfo {author} {\bibfnamefont {R.}~\bibnamefont
  {Olfati-Saber}}, \bibinfo {author} {\bibfnamefont {J.~A.}\ \bibnamefont
  {Fax}},\ and\ \bibinfo {author} {\bibfnamefont {R.~M.}\ \bibnamefont
  {Murray}},\ }\href {https://doi.org/10.1109/JPROC.2006.887293} {\bibfield
  {journal} {\bibinfo  {journal} {Proc. IEEE}\ }\textbf {\bibinfo {volume}
  {95}},\ \bibinfo {pages} {215} (\bibinfo {year} {2007})}\BibitemShut
  {NoStop}%
\bibitem [{\citenamefont {FitzHugh}(1961)}]{FitzHugh1961}%
  \BibitemOpen
  \bibfield  {author} {\bibinfo {author} {\bibfnamefont {R.}~\bibnamefont
  {FitzHugh}},\ }\href {https://doi.org/10.1016/S0006-3495(61)86902-6}
  {\bibfield  {journal} {\bibinfo  {journal} {Biophys. J.}\ }\textbf {\bibinfo
  {volume} {1}},\ \bibinfo {pages} {445} (\bibinfo {year} {1961})}\BibitemShut
  {NoStop}%
\bibitem [{\citenamefont {van~den Driessche}\ and\ \citenamefont
  {Watmough}(2002)}]{vandenDriessche2002}%
  \BibitemOpen
  \bibfield  {author} {\bibinfo {author} {\bibfnamefont {P.}~\bibnamefont
  {van~den Driessche}}\ and\ \bibinfo {author} {\bibfnamefont {J.}~\bibnamefont
  {Watmough}},\ }\href {https://doi.org/10.1016/S0025-5564(02)00108-6}
  {\bibfield  {journal} {\bibinfo  {journal} {Math. Biosci.}\ }\textbf
  {\bibinfo {volume} {180}},\ \bibinfo {pages} {29} (\bibinfo {year}
  {2002})}\BibitemShut {NoStop}%
\bibitem [{\citenamefont {Trefethen}\ and\ \citenamefont
  {Embree}(2005)}]{Trefethen2005}%
  \BibitemOpen
  \bibfield  {author} {\bibinfo {author} {\bibfnamefont {L.~N.}\ \bibnamefont
  {Trefethen}}\ and\ \bibinfo {author} {\bibfnamefont {M.}~\bibnamefont
  {Embree}},\ }\href@noop {} {\emph {\bibinfo {title} {Spectra and
  Pseudospectra: The Behavior of Nonnormal Matrices and Operators}}}\ (\bibinfo
   {publisher} {Princeton University Press},\ \bibinfo {address} {Princeton,
  NJ},\ \bibinfo {year} {2005})\BibitemShut {NoStop}%
\bibitem [{\citenamefont {Hairer}\ and\ \citenamefont
  {Wanner}(1996)}]{HairerWanner1996}%
  \BibitemOpen
  \bibfield  {author} {\bibinfo {author} {\bibfnamefont {E.}~\bibnamefont
  {Hairer}}\ and\ \bibinfo {author} {\bibfnamefont {G.}~\bibnamefont
  {Wanner}},\ }\href {https://doi.org/10.1007/978-3-662-09947-6} {\emph
  {\bibinfo {title} {Solving Ordinary Differential Equations {II}: Stiff and
  Differential-Algebraic Problems}}},\ \bibinfo {edition} {2nd}\ ed.\ (\bibinfo
   {publisher} {Springer},\ \bibinfo {address} {Berlin},\ \bibinfo {year}
  {1996})\BibitemShut {NoStop}%
\bibitem [{\citenamefont {Holme}\ and\ \citenamefont
  {Saram{\"a}ki}(2012)}]{Holme2012}%
  \BibitemOpen
  \bibfield  {author} {\bibinfo {author} {\bibfnamefont {P.}~\bibnamefont
  {Holme}}\ and\ \bibinfo {author} {\bibfnamefont {J.}~\bibnamefont
  {Saram{\"a}ki}},\ }\href {https://doi.org/10.1016/j.physrep.2012.03.001}
  {\bibfield  {journal} {\bibinfo  {journal} {Phys. Rep.}\ }\textbf {\bibinfo
  {volume} {519}},\ \bibinfo {pages} {97} (\bibinfo {year} {2012})}\BibitemShut
  {NoStop}%
\bibitem [{\citenamefont {Kohler}\ \emph {et~al.}(2005)\citenamefont {Kohler},
  \citenamefont {Lehmann},\ and\ \citenamefont {H{\"a}nggi}}]{floquet2005}%
  \BibitemOpen
  \bibfield  {author} {\bibinfo {author} {\bibfnamefont {S.}~\bibnamefont
  {Kohler}}, \bibinfo {author} {\bibfnamefont {J.}~\bibnamefont {Lehmann}},\
  and\ \bibinfo {author} {\bibfnamefont {P.}~\bibnamefont {H{\"a}nggi}},\
  }\href {https://doi.org/10.1016/j.physrep.2004.11.002} {\bibfield  {journal}
  {\bibinfo  {journal} {Phys. Rep.}\ }\textbf {\bibinfo {volume} {406}},\
  \bibinfo {pages} {379} (\bibinfo {year} {2005})}\BibitemShut {NoStop}%
\bibitem [{\citenamefont {Coletta}\ and\ \citenamefont
  {Jacquod}(2016)}]{coletta2016}%
  \BibitemOpen
  \bibfield  {author} {\bibinfo {author} {\bibfnamefont {T.}~\bibnamefont
  {Coletta}}\ and\ \bibinfo {author} {\bibfnamefont {P.}~\bibnamefont
  {Jacquod}},\ }\href {https://doi.org/10.1103/PhysRevE.93.032222} {\bibfield
  {journal} {\bibinfo  {journal} {Phys. Rev. E}\ }\textbf {\bibinfo {volume}
  {93}},\ \bibinfo {pages} {032222} (\bibinfo {year} {2016})}\BibitemShut
  {NoStop}%
\bibitem [{\citenamefont {Tang}\ and\ \citenamefont
  {Bassett}(2018)}]{tang2018}%
  \BibitemOpen
  \bibfield  {author} {\bibinfo {author} {\bibfnamefont {E.}~\bibnamefont
  {Tang}}\ and\ \bibinfo {author} {\bibfnamefont {D.~S.}\ \bibnamefont
  {Bassett}},\ }\href {https://doi.org/10.1103/RevModPhys.90.031003} {\bibfield
   {journal} {\bibinfo  {journal} {Rev. Mod. Phys.}\ }\textbf {\bibinfo
  {volume} {90}},\ \bibinfo {pages} {031003} (\bibinfo {year}
  {2018})}\BibitemShut {NoStop}%
\bibitem [{\citenamefont {Wang}(2026)}]{wang2026centrality}%
  \BibitemOpen
  \bibfield  {author} {\bibinfo {author} {\bibfnamefont {Y.}~\bibnamefont
  {Wang}},\ }\href {https://doi.org/10.5281/zenodo.22792990} {\bibinfo {title}
  {Multi-dimensional dynamical centrality from green functions in complex
  networks}} (\bibinfo {year} {2026}),\ \bibinfo {note}
  {https://doi.org/10.5281/zenodo.22792990}\BibitemShut {NoStop}%
\end{thebibliography}%

\clearpage

\end{document}